\documentclass[traditabstract]{aa-package/aa}

\usepackage[varg]{txfonts}
\usepackage{time}
\usepackage{color}
\usepackage{natbib}
\usepackage[percent]{overpic}
\usepackage{amstext,amsmath,mathtools}
\usepackage{graphicx}
\usepackage{textcomp}
\usepackage{txfonts}
\usepackage{url}
\usepackage{booktabs} 

\usepackage[labelformat=simple]{subcaption}
\renewcommand\thesubfigure{(\alph{subfigure})}

\newcommand{\gs}{}
\newcommand{\gss}{}

\DeclareMathOperator{\cov}{cov}
\DeclareMathOperator{\diag}{diag}

\def\ip#1#2{\left\langle#1\;\middle|\;#2\right\rangle}

\def\gcmatrix{\mathbf{\Delta}}           
\def\acelement{C}                        
\def\acmatrix{\mathbf{\acelement}}       
\def\cmmatrix{\boldsymbol{\mathfrak{B}}} 
\def\mmmatrix{\mathbf{M}}                
\def\ifelement{G}                        
\def\ifmatrix{\mathbf{\ifelement}}       
\def\dmmatrix{\mathbf{A}}                
\def\esmatrix{\mathbf{D}}                

\begin{document}
\titlerunning{The SST AO system}
\title{The 85-electrode AO system of the Swedish 1-m Solar Telescope}

\author{G.B. Scharmer\inst{1,2,3}
\and
G. Sliepen\inst{1,2}
\and 
J.-C. Sinquin\inst{4}
\and
M. G. L\"ofdahl\inst{1,2}
\and
B. Lindberg\inst{5}
\and 
P. S\"utterlin\inst{1,2}}

\institute{Institute for Solar Physics, Stockholm University,
AlbaNova University Center, SE 106\,91 Stockholm, Sweden \and
Stockholm Observatory, Dept. of Astronomy, Stockholm University,
AlbaNova University Center, SE 106\,91 Stockholm, Sweden \and
Royal Swedish Academy of Sciences, Box 50005, SE 104\,05 Stockholm, Sweden \and 
CILAS, 8 Avenue Buffon, CS 16319, 45063 Orléans Cedex 2 - France \and
Lens Tech AB, Tallbackagatan 11, SE 931\,64 Skellefte\aa, Sweden}
\date{Draft: \now\ \today}
\frenchspacing

\abstract{We discuss the {\gss chosen concepts, detailed design, implementation and calibration of the 85-electrode adaptive optics (AO) system of the Swedish 1-meter Solar Telescope (SST), which was installed in 2013. The AO system is unusual by using a combination of a monomorph mirror with a Shack-Hartmann (SH) wavefront sensor (WFS), and by using a second high-resolution SH microlens array to aid the DM characterization, calibration, and modal control. An Intel PC workstation performs the heavy image processing associated with cross correlations and real-time control at 2~kHz update rate with very low latency.} The computer and software continue the successful implementation since 1995 of earlier generations of correlation tracker and AO systems at SST and its predecessor SVST by relying entirely on work station technology and an extremely efficient algorithm for implementing cross correlations with the large field-of-view of the WFS. We describe critical aspects of the design, calibrations, software and functioning of the AO system. The exceptionally high performance is testified through the highest Strehl ratio {\gss (inferred from the measured granulation contrast) of existing meter-class solar telescopes, as demonstrated here at wavelengths shorter than 400~nm and discussed in more detail in a separate publication by Scharmer et al.} We expect that some aspects of this AO system may be of interest also outside the solar community.

}

\maketitle

\section{Introduction}
Achieving near diffraction limited spatial resolution with high Strehl at visible wavelengths is of paramount importance for major ground-based solar telescopes such as the Daniel K. Inouye Solar Telescope \citep[DKIST;][]{2008AdSpR..42...78R,2017psio.confE..79M}, which is now in full operation in Maui \citep{2018SPIE10700E..0VW,2020SoPh..295..172R}, and the future European Solar Telescope \citep[EST;][]{2013MmSAI..84..379C,2016SPIE.9908E..09M,2019AdSpR..63.1389J, 2022A&A...666A..21Q}, which is still in its design stage. Meanwhile, major existing meter-class solar telescopes such as the Swedish 1-meter Solar Telescope \citep[SST;][]{2003SPIE.4853..341S} on La Palma, the Goode Solar Telescope \citep[GST;][]{2010AN....331..620G} at Big Bear Solar Observatory and GREGOR \citep{2012ASPC..463..365S} on Tenerife, are used by the solar community to achieve some of the scientific goals of DKIST and EST, and to constitute platforms for the development of instrumentation for existing major and future solar telescopes. The performance of all these telescopes depends critically on the successful implementation of powerful adaptive optics (AO) systems to feed spectroscopic and polarimetric instrumentation, which are key to advancing the frontier of solar physics. 

A particular challenge for solar telescopes is that solar fine structure almost always is observed against a luminous background. Observational diagnostics of dynamic and magnetic solar fine structure is therefore not simply a question of resolving structures of the smallest scales, but even more so of observing such structures at the highest possible contrast against the background. This calls for observations at the highest possible Strehl. Such observations at visible wavelengths are in turn made very difficult by the relatively poor ground-layer seeing during daytime, which often is not better than 1\farcs0, even at the best sites \citep{2006SPIE.6267E..1TH,2011LRSP....8....2R}\footnote{Rough estimates for La Palma suggest that the fraction of time with better than 1" daytime seeing (measured along the line of sight to the Sun) is in the range 1--10\%, depending on the height of the telescope above the ground, and to what extent high-altitude seeing from around the tropopause is included in the measurements.}. Achieving diffraction limited resolution at high Strehl in such seeing conditions requires a wavefront sensor (WFS) with small subapertures \citep[on the order of 10~cm, or less - see articles by F. Roddier in][]{2004aoa..book.....R} and a deformable mirror matched to the WFS. This provides another challenge: that the most easily accessible WFS target on the solar surface, which is the solar granulation pattern, has low contrast and spatial scales of about 1\farcs5 \citep{1967sogr.book.....B,1969SoPh....7..167N}. Reducing the WFS subaperture size gradually below about 10~cm, therefore leads to a steep increase of the wavefront sensor noise. This limits subaperture diameters for solar AO to about 8~cm or more \citep{2010AN....331..640B}, though in practice 9-10~cm perhaps is a safer choice. Furthermore, the use of a poorly resolved random pattern of granules as wavefront sensor target requires a wavefront sensor with relatively large field-of-view to provide stable positional information with acceptable noise, and therefore the employment of large format wavefront sensor cameras and compute-intensive cross-correlation algorithms. 

Solar AO thus shares common features with night-time AO, such as the use of Shack-Hartmann wavefront sensors. However, the use of low-contrast solar granulation pattern as WFS target, in combination with the relatively poor daytime seeing, presents particular challenges that are key in driving the development of solar AO.
Such solar AO systems developed up to about 2010, have been reviewed by \citet{1999aoa..book..235B}, \citet{2004SPIE.5490...34R}, and \citet{2011LRSP....8....2R}, the latter with particular emphasis on the the AO76 system developed for the R.B. Dunn Solar Telescope \citep{2004SPIE.5490...34R}, and later by \citet{2016SPIE.9909E..0XS}. Other recently developed major AO systems for solar telescopes involve that for GREGOR \citep{2012AN....333..863B,2016SPIE.9909E..24B,2018SPIE10703E..3AB}, for GST \citep{2014SPIE.9148E..35S,2014SPIE.9148E..2US}, for DKIST \citep{2020SPIE11448E..0TJ} and for the Chinese New Vacuum Solar Telescope \citep{2023SCPMA..6669611Z}. 

The SST 85-electrode AO system is different from other solar AO systems in several respect. A first brief description of this AO system can be found in \citet{2014SPIE.9148E..0GS}, and an exhaustive analysis of the image quality obtained with the AO system and the SST science instrumentation can be found in \citet{2019A&A...626A..55S}. Here, the constraints and chosen concepts, detailed design, implementation and calibration of the AO system are presented for the first time. Section 2 gives an overview of the concepts and constraints of the present design and of the earlier AO systems developed for SST, then describes the deformable mirror (DM), the SH WFS, and the control electronics. In Sect.~3, we describe the control computer, the software and calibrations, and in Sect.~4 we comment on the performance. In Sect.~5, finally, we summarize and end with a few concluding remarks of a more general character.

\begin{figure}
\center
\includegraphics[angle=0, width=\linewidth,clip]{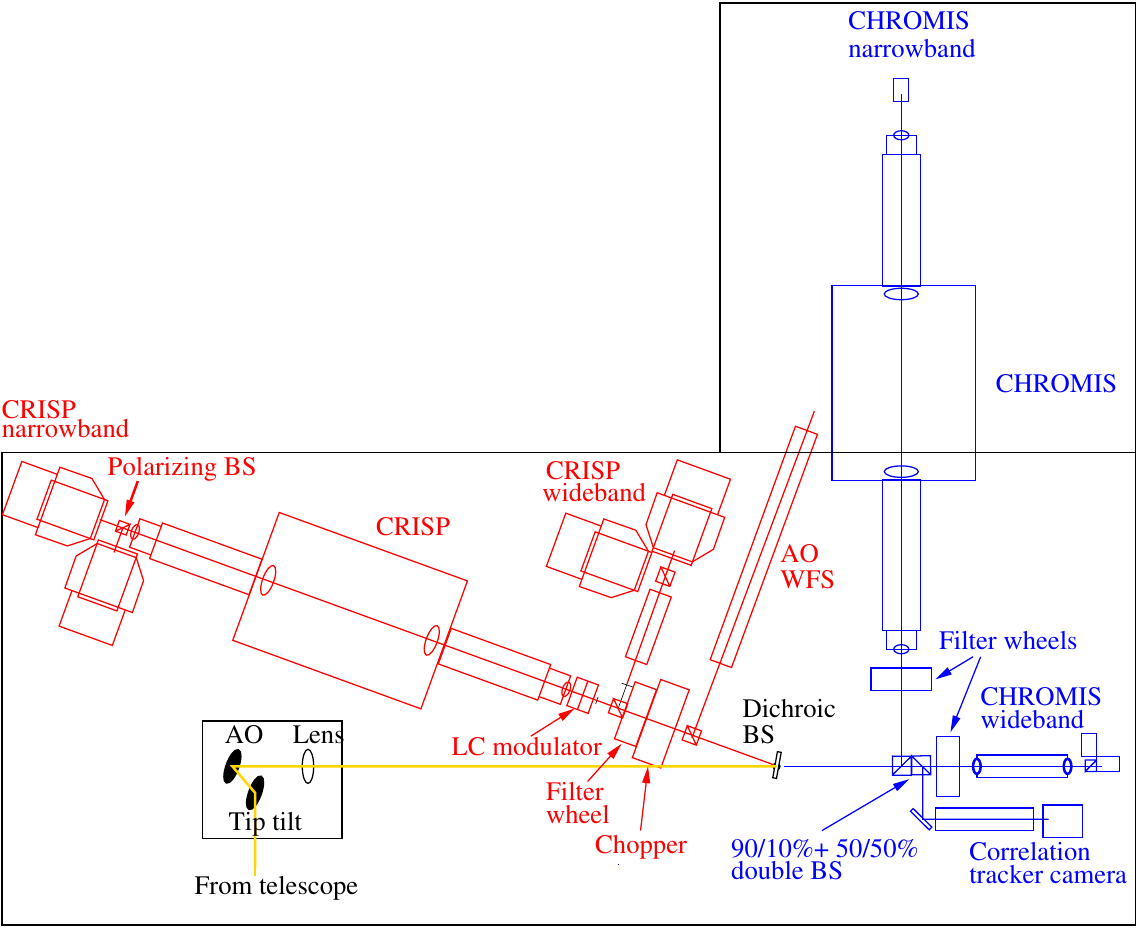}
 \caption{
Layout of the optical setup at SST, including the tip-tilt and adaptive mirrors and the wavefront sensor. The CRISP and CHROMIS narrowband and wideband re-imaging systems are also shown. Recently, one more science instrument (HeSP), has been installed, which is not shown in the drawing, adjacent to CHROMIS. 
}
\label{fig:Re-imaging}
\end{figure}

\section{Wavefront sensor, deformable mirror and functioning of the AO system}
\subsection{Background and conceptual design} \label{sect:background}


The first AO system for SST was developed at the 50-cm Swedish Vacuum Solar Telescope \citep[SVST;][]{1985ApOpt..24.2558S}, which became the first European solar telescope equipped with AO \citep{2000SPIE.4007..239S}. However, this was only a temporary installation - the successful deployment of this AO system on SVST merely served as a demonstrator, and was the prerequisite for the decision to construct the 1-meter SST. This AO system was thus integrated in optical the design of SST, such that SST became the first major solar telescope with an AO system that was an integral part of its design. This in turn defined certain constraints, in particular the AO pupil diameter, for future generations of AO systems on SST. It can be remarked that the inclusion of the AO system in the optical design  of SST only introduces two extra optical components in the science beam: the first is one extra mirror (the deformable mirror itself), the second is a 5-10~\% beamsplitter cube close to the science focus of the F/46 beam. Overall, the optical system of SST is designed to be as simple as possible to minimize image degradation and to maximize throughput, and of the highest possible optical quality.

Based on the impressive performance of the 19-electrode/subaperture AO system PUEO on the Canada-France-Hawaii Telescope \citep{1998PASP..110..152R}, which is based on Roddier's curvature concept \citep{1991SPIE.1542..248R, 2004aoa..book.....R}, the decision was taken by one of the authors (Scharmer) to adopt a similar bimorph mirror to a 19-subaperture WFS with hexagonal subapertures. {\gs Although the high efficiency of such low-order bimorph mirrors had then only been demonstrated in combination with wavefront curvature sensors \citep{1998PASP..110..837R}, it was conjectured that high efficiency could also be obtained by combining a low-order bimorph mirror with a Shack-Hartmann {\gs (SH) wavefront sensor (WFS)} that is well aligned with the electrode layout. This conclusion was based on the fact that bimorph mirrors deliver a spatial spectrum of wavefront displacements that is close to the spectrum of wavefront distortions from turbulent seeing \citep{1992aolt.meet...59R}. Since turbulent seeing is best characterized by Karhunen--Loeve (KL) aberrations, with small amplitudes for the high-order aberrations that makes the wavefronts "smooth", and since the most efficient way of compensating  for seeing with a given number of degrees of freedom is by removing the lowest-order KL aberrations \citep{2004aoa..book.....R}, low-order bimorph mirrors can in principle provide close to optimum wavefront compensation of seeing with a given number of degrees of freedom \citep{1991PASP..103..131R}. This "smoothness" of wavefronts from bimorph mirrors (which is also a characteristic of the so-called monomorph mirrors used in the present AO system - see Sect.~\ref{sect:architecture}) moreover suggests that they should work well in combination with a SH WFS, and this solution was therefore judged to constitute a promising choice for SST (see also Sect. \ref{sect:architecture}).}

{\gs The pupil diameter of the bimorph mirror was decided in consultation with the manufacturer, AOPTIX Technologies Inc. (formerly Laplacian Optics), to be 34~mm. This was considered the smallest possible pupil diameter that would allow future upgrades of the SST AO system compensating up to about 60 KL modes, which was at that time foreseen to be the upper limit for future developments \citep{2003SPIE.4853..370S}. Keeping the pupil diameter small, on the other hand, was considered important in order to make the associated re-imaging system compact (to fit within the available space without the use of folding mirrors), to make the re-imaging system less sensitive to internal seeing, to maximize the resonance frequencies of the bimorph mirror, and to minimize challenges and costs of  manufacturing a high-quality mirror.}

It was also decided to compensate atmospheric image motion with a separate tip-tilt mirror, which is controlled by a separate correlation tracker camera and computer. {\gs There are several reasons for preferring this arrangement. One argument is purely optical: that folding the beam from vertical to horizontal using only the bimorph mirror would require the pupil to be strongly elliptical ($34\times 48$~mm), thus making its manufacture more complex and its performance likely less satisfactory. By folding the beam first 60\degr{} with the tip-tilt mirror, and then only 30\degr{} with the bimorph mirror, we obtain a pupil that is almost circular (since cos 15\textdegree=0.97) on the downward-facing (which provides some protection against dust) reflective surface of the bimorph. The other arguments for using a separate tip-tilt mirror are related to the importance of stable image quality and user friendliness of the entire system. During periods of poor seeing, the AO system automatically goes into hibernation and then recovers by itself when the seeing gets better, and does not need attention from the observer even when the pointing of the telescope is changed. The telescope thus consistently delivers the best possible image quality, while also minimizing the needed interactions by the observer. Moreover, the correlation tracker of the tip-tilt mirror can use a much larger field of view than is possible with the wavefront sensor of the AO system, making it possible to maintain tracking through periods of very poor seeing. For solar observations, this is important because day-time seeing is often extremely intermittent. Overall, these are highly appreciated features of the AO and correlation tracker systems. Finally, since the adaptive mirror does not need to compensate for the tip-tilt modes, more of its available stroke is available for correcting higher-order modes.}

The SST re-imaging system was therefore designed with a singlet lens acting as field lens and exit vacuum window, and delivering a 34-mm pupil image at the given height above the optical table on which science instrumentation and cameras are mounted. It should be remarked, that increasing the pupil diameter is possible but not simple, and is unlikely to ever happen: this would require lowering the height of the aforementioned optical table and replacing the field lens, the re-imaging triplet lens, and requiring a major re-arrangement of the optical tables with their science instrumentation.

As regards the design of the wavefront sensor, the physical diameter of the 85-subaperture microlens array is given by the size of the sensor of the chosen WFS camera, and its focal length is set by the chosen image scale of 0\farcs48 and the f-ratio (F/46) of the re-imaging beam at the focal plane of the WFS. A novelty of the 85-electrode AO system is the use of a second microlens array that has the same focal length as the first array but 253 useful subapertures within the pupil. This is used for characterization of the electrode responses with better pupil resolution than is possible with the 85-subaperture array, as described in Sect. \ref{sec:establ-contr-matr}. 

A final decision concerned the importance of constraining the AO system development costs, and this was to build the system on work station technology. The proof of concept of this strategy was the development of a correlation tracker system installed on SVST in August 1994 \citep{1995SPIE.2607..145S}. To enable the use of this technology, the time consuming cross-correlations were implemented with an algorithm developed {\gs in 1993} by one of the authors (Scharmer), usually referred to as the absolute difference squared (ADF$^2$) algorithm, which allowed the accurate and efficient use of multi-media instructions \citep[in particular the pixel error instruction -- PERR;][]{1999ASPC..183..231S}. The accuracy of the ADF$^2$ algorithm was compared to that of other algorithms by \citet{2010A&A...524A..90L} and found, as expected, to be only marginally less accurate than the square difference  algorithm but superior to algorithms based on Fast Fourier Transforms\footnote{This conclusion has recently been contested by \citet{Wei_2023}, who conclude that the so-called Covariance Function in the frequency domain (CFF) provides superior measurement accuracy and robustness. {\gs However, their analysis is peculiar in that this superior accuracy seems to stem from pre-processing of sunspot test data, which is applied only to the CFF and one other method, but not to remaining methods, such as the ADF$^2$ algorithm, (cf. their Sect.~3.3)}}. The  ADF$^2$  algorithm was implemented in the first and second generation AO system for SST, the latter of which was based on a 37-electrode bimorph mirror from AOPTIX \citep{2003SPIE.4853..370S}, and also in the present 85-electrode system.

\subsection{SST optical system} \label{sect: SST_optics}

The SST \citep{2003SPIE.4853..341S} is a 1-meter evacuated telescope with an optical system that can be divided into a primary optical system (the 1.1~m fused silica singlet lens, which is also the entrance vacuum window, and two 1.4~m Zerodur mirrors used at 45\degr{} angle of incidence) and a secondary optical system. The latter consists of a 60~mm field mirror, a 250~mm clear aperture so-called Schupmann corrector (consisting of a meniscus fused silica lens used in double pass and a concave Zerodur mirror), a field lens (which is also the exit vacuum window) that produces a pupil image with 34~mm diameter, a tip-tilt mirror, a DM and a triplet lens that produces an F/46 beam at the optical tables. The beam exiting the vacuum system above the optical table is nearly vertical and is folded twice: first by a 42~mm tip-tilt mirror that is used at an angle of incidence of 30\degr{} and then by the DM, which is used at an angle of incidence of 15\degr. This results in a horizontal beam but requires the DM to have a slightly elliptic shape. Figure~\ref{fig:Re-imaging} shows the layout of the SST main optical table, with the tip-tilt mirror and DM (marked as ``AO''), and the wavefront sensor (WFS) indicated.  We note that the impact of the AO system on the science beam is very small -- it adds only one mirror plus a 5--10\% beamsplitter cube close to the science focal plane. We also note that the SST has a perfectly round aperture without any (central) obscuration.

\subsection{Microlens layout}

\begin{figure}[!t]
\center
\includegraphics[angle=0, width=0.95\linewidth,clip]{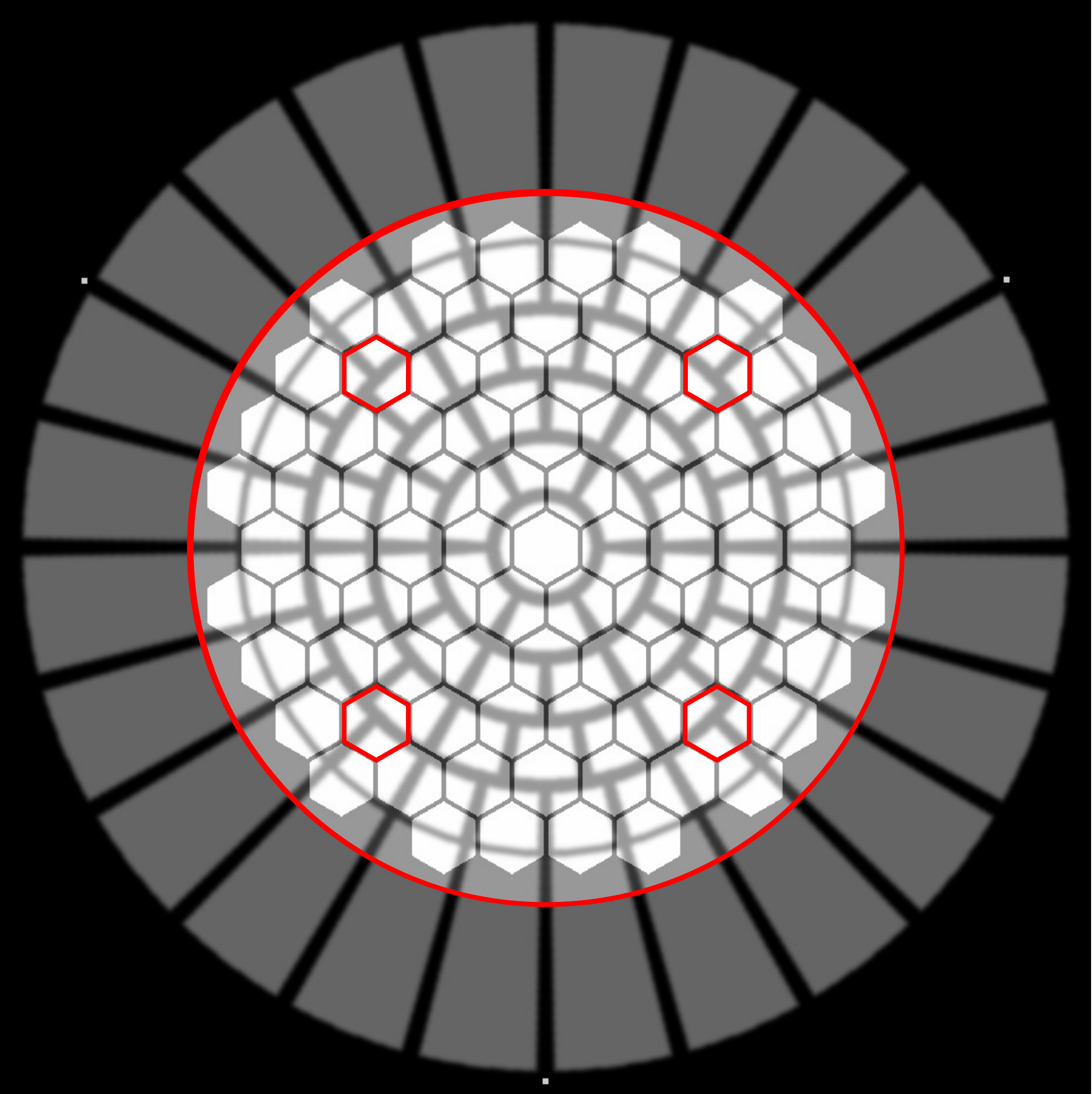}
\includegraphics[angle=0, width=0.95\linewidth,clip]{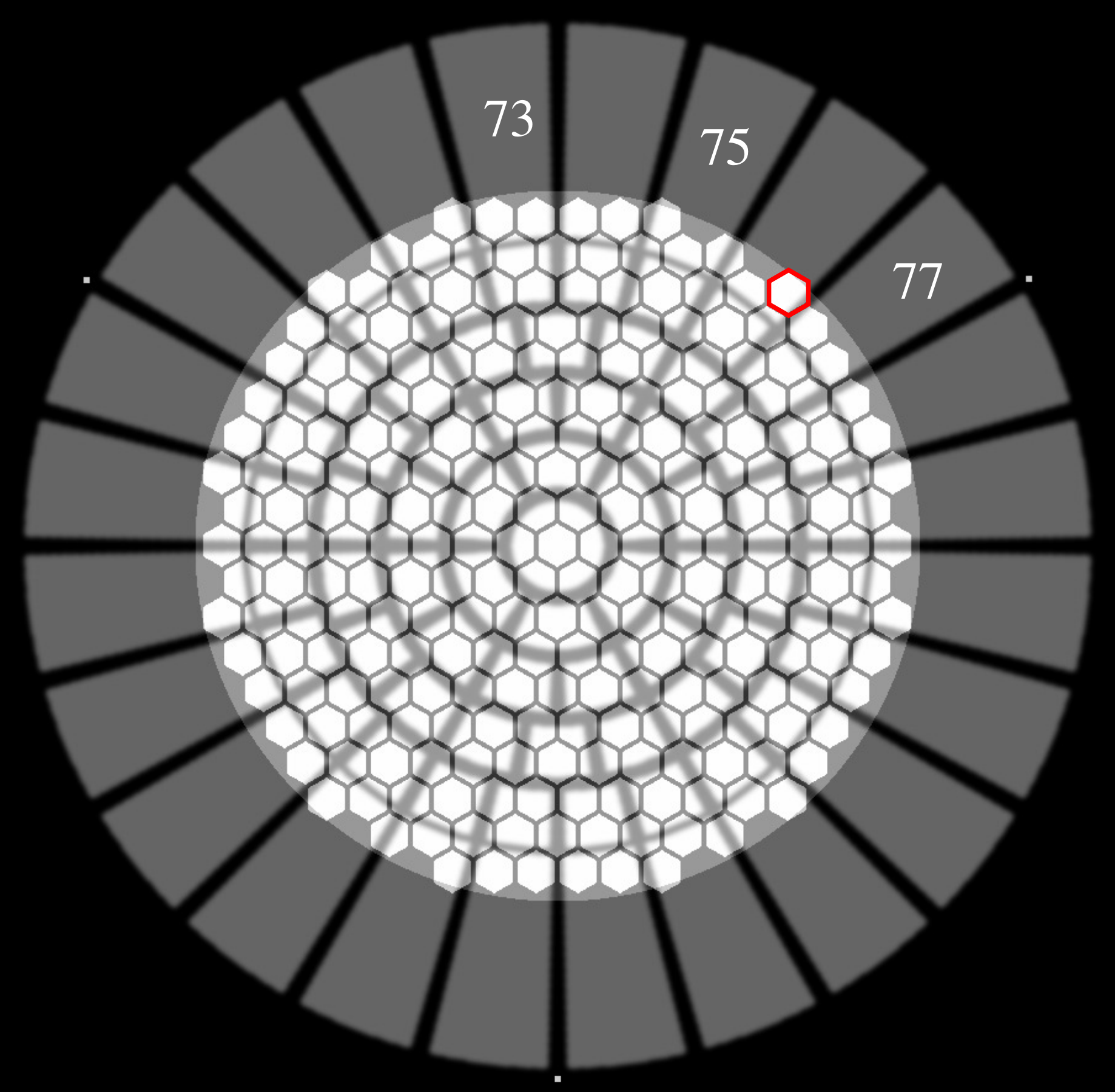}
 \caption{
The upper panel shows the layout of the 85-subaperture wavefront sensor and the 85-electrode monomorph DM of the SST AO system. The gray or white radial/azimuthal structures correspond to the electrodes of the DM, and the hexagonal structures correspond to the 85 lenslets. The large red circle corresponds to the pupil diameter and the four lenslets highlighted in red correspond to the subapertures used for seeing measurements. The lower panel shows the layout of the 253-subaperture WFS, used to calibrate the wavefronts (influence functions) produced by the 85 electrodes of the DM. Electrodes 73, 75 and 77, and their relation to microlens 16 (highlighted in read), for which electrode responses are shown in Fig.~\ref{fig:DM_response2}, are indicated in the lower panel. 
}
\label{fig:AO_WFS}
\end{figure}

The SST AO system uses solar granulation, with typical scales of 1\farcs5, and other solar fine structure as WFS target. Solar granulation observed at low spatial resolution has very low contrast, which imposes constraints on the angular resolution of the WFS. The AO76 system designed for the R.B. Dunn Solar Telescope \citep{2004SPIE.5490...34R}, also discussed in detail by \citet{2011LRSP....8....2R}, uses square subapertures with a diameter of 7.5~cm, consistent with the conclusion of \citet{2010SPIE.7736E..2JB} that a subaperture diameter of about 8~cm represents a lower limit for obtaining granulation contrast of a few per cent. For SST, we use hexagonal microlenses with larger apertures of about 9.4~cm, after having concluded that a WFS with 85 hexagonal subapertures could provide an adequate coverage of the 98~cm clear aperture of SST, and that the 9.4~cm subapertures would provide an acceptable spatial resolution of the granulation pattern. The layout of the WFS is shown in Fig.~\ref{fig:AO_WFS}, with the pupil diameter indicated as a red circle in the upper panel. 

\subsection{Deformable mirror}

\subsubsection{Architecture and main characteristics}
\label{sect:architecture}

After considerable discussion, it was concluded that the DM should be an 85-electrode so-called monomorph mirror, manufactured by CILAS but with an electrode geometry designed by one of the authors (Scharmer), as described in Sect. \ref{sect:electrode_layout}. {\gs As pointed out in Sect. \ref{sect:background}, bimorph and monomorph mirrors share the desirable property that they produce "smooth" wavefronts.} Specifically, the spatial spectrum of bending deformations decreases as $k^{-2}$, where $k$ is the spatial wavenumber, which is close to the $k^{-11/6}$ decrease of the Kolmogorov spectrum of phase fluctuations \citep{1992aolt.meet...59R}. This reduces the production of a ``tail'' of unwanted high-order aberrations when compensating lower-order aberrations with the AO system.  

\begin{figure}
\includegraphics[angle=0, width=0.48\textwidth,clip]{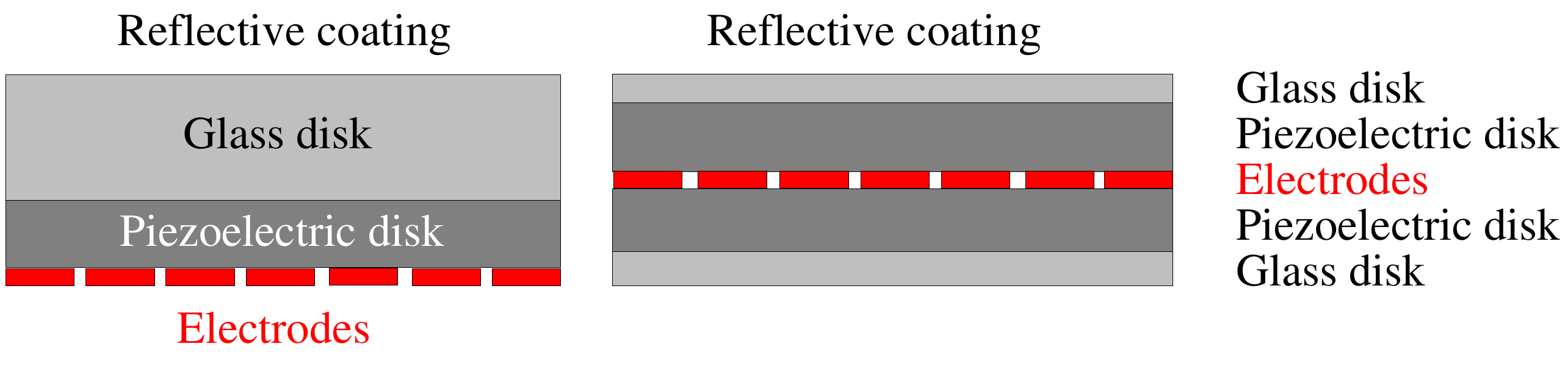}
 \caption{
Schematic drawings illustrating the different architectures of a monomorph (left) and bimorph (right) DM. 
}
\label{fig:dm_architectures}
\end{figure}

As illustrated schematically in Fig.~\ref{fig:dm_architectures}, a monomorph DM is made of two main components bonded together: a piezo plate and a glass plate that is polished and covered with a reflective coating. The piezo plate is equipped with many (in our case, 85) electrodes. When a voltage is applied on an electrode, the electrical field induces a local in-plane expansion or contraction of the piezo plate. The bimetallic effect between the piezoelectric ceramics and the optical plate then generates a local curvature of the optical surface. The optical surface of the mirror is controlled by applying voltage on each electrode to generate the desired deformation within the pupil.

\begin{figure}
\center
\includegraphics[width=\linewidth,clip]{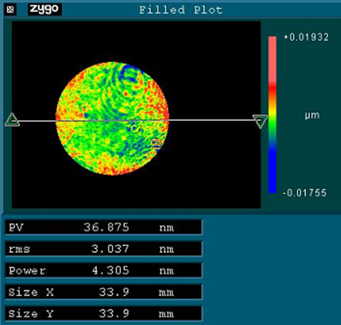}
 \caption{
Optical quality inside the 34~mm pupil of one of the two DMs delivered, after flattening the mirror with optimum voltages on the 85 electrodes. The RMS surface error is 3~nm (wavefront 6~nm RMS). There are no discernible print-through effects from the electrode pattern.   
}
\label{fig:DM_quality}
\end{figure}

A bimorph mirror uses the same functional concept as a monomorph mirror but its architecture is more complex. A bimorph mirror is made of two piezo plates sandwiched in between two glass plates, as shown in Fig.~\ref{fig:dm_architectures}. The local curvatures of the optical surface are generated by the expansion of one of the two piezo plates, and contraction of the other. Since the electrodes are located at the interface between the two piezo plates, the electrical contacting is critical. In addition, this architecture requires very thin glass plates to achieve sufficient stroke. The optical plates of a bimorph mirror needs to be ten times thinner than that of a monomorph mirror to generate the same stroke. As a result, a monomorph mirror has electrodes and electrical contacting that are further away from the optical surface than for a bimorph mirror, which avoids any print-through effect from the electrode pattern. This allows polishing that can reach an excellent optical quality after flattening -- for the present DM that quality is estimated at 6~nm RMS wavefront error, as shown in Fig.~\ref{fig:DM_quality}. 
This high optical quality is of considerable importance since the SST AO system is used for observations at wavelengths shorter than 400~nm. 

A challenge of the design of the DM was to make the glass plate onto which the piezo plate was glued sufficiently thin to achieve the required stroke and sufficiently thick to achieve a first resonance frequency of at least 2.7~kHz (goal 3.7~kHz). The 52~mm mirror was mounted at 3 points along its outer edge in a 100~mm diameter, 80~mm thick aluminum cell and used 3 additional dampers to suppress the lowest resonance modes. 

\subsubsection{Electrode layout}\label{sect:electrode_layout}

Based on earlier experience with the design of the 37-electrode bimorph mirror and wavefront sensor for SST \citep{2003SPIE.4853..370S}, the DM was designed using software developed by one of the authors (Scharmer) and with the following constraints:
\begin{itemize}
 \item to be for a pupil diameter of 34~mm, stretched slightly into an elliptic shape of 35.2$\times$34~mm by the 15\degr{} angle of incidence
 \item to have (a maximum of) 85 electrodes in order to match the number of degrees of freedom of the WFS
 \item to have 0.7 mm gaps between the electrodes (to eliminate the risk of  short circuits)
 \item to match the layout of the microlens array as well as possible, meaning that the electrodes should be located away from microlens centers and if possible be such that the microlenses cover junctions of two or more electrodes
 \item that all electrodes, except those outside the pupil, should have roughly the same area to deliver similar stroke (require similar voltage) irrespective of the location of the electrode within the pupil
 \item to provide excellent compensation for low-order KL and Zernike aberrations, such as focus and astigmatism
 \item to provide the best possible wavefront compensation of turbulent seeing induced aberrations, assumed to obey Kolmogorov statistics
 \item to minimize the diameter of the outermost electrode ring in order to maintain a small overall mirror diameter such that the DM resonance frequency can be maximized
\end{itemize}

{\gs Initially, a few designs with different number of electrode rings were tried, but soon focus was on designs with a center electrode and four surrounding rings of electrodes. The number of electrode rings and azimuth angles were chosen to match the radial and azimuthal orders of roughly the 50 first KL polynomials, corresponding to the layout shown in Fig.~\ref{fig:AO_WFS}. Initially, the radii of the electrode rings were adjusted manually and the matching of the electrode geometry to the microlens layout was evaluated with respect to both the expected residual wavefront variance. However, rather different radii delivered very small differences in residual wavefront variance. 
Instead, the condition number (calculated as the ratio of the smallest to largest singular values) was used to evaluate and optimize the match between the electrode layout and the SH WFS geometry. This corresponds to } SVD (singular value decomposition -- see below) of the matrix $\mathbf{C}$, defined through the relations
\begin{align}
 \mathbf{x} &= \mathbf{A}\cdot \mathbf{k},\\
 \mathbf{k} &= \mathbf{B}\cdot \mathbf{v},\\ 
 \noalign{\noindent and}
 \mathbf{x} &= (\mathbf{A}\cdot \mathbf{B})\cdot \mathbf{v}=\mathbf{C}\cdot \mathbf{v},
\end{align}
where $\mathbf{x}$ corresponds to the measured $(x,y)$ subimage positions, $ \mathbf{k}$ the KL coefficients, $\mathbf{A}=\mathbf{A}(2N_{S},N_K)$ represents the connection between these two quantities 
and is obtained theoretically, and $\mathbf{v}$ the voltages applied to the DM electrodes. For our simulations, we used $N_{K}=400$ modes. The matrix $\mathbf{B}$ contains the $N_K$ KL coefficients from each of the $N_e$ electrodes and is obtained from thin-plate theory, by modeling the response of the mirror surface to applied voltages as the solution to the Poisson equation \citep{Kokorowski:79,1988ApOpt..27.1223R,1998aoat.book.....H}, where the right-hand side of the equation is unity inside a specified electrode and zero outside. The equation was solved with Fourier methods, taking into account the 3-point mounting of the mirror shown in Fig.~\ref{fig:AO_WFS}, but not the 3 dampers in between the mounting points, and eliminating wrap-around effects from the use of Fourier transforms. The matrix $\mathbf{C}$ thus corresponds to the interaction matrix (or ``poke'' matrix), which is calibrated routinely for the installed AO system at SST. This matrix can be inverted with SVD methods and the so-obtained singular values constitute a good measure of the quality of the match between the SH WFS and the electrode geometry. The pseudo-inverse of $\mathbf{C}$ can furthermore be combined with $\mathbf{A}$ to obtain the reconstruction matrix $\mathbf{E}$,
\begin{equation}
 \mathbf{v}= \mathbf{C}^\text{I}\cdot \mathbf{x}=(\mathbf{C}^\text{I}\cdot\mathbf{A})\cdot\mathbf{k}=\mathbf{E}\cdot\mathbf{k}.
\end{equation}
We used this equation to make a simple closed-loop simulation for estimating the performance of various electrode layouts. For the layout finally chosen, this gave the following predictions:
\begin{itemize}
 \item number of correctable modes: 84
 \item residual wavefront variance normalized to the input variance: 0.0074 if tip-tilt modes are included, 0.054 if tip-tilt modes are excluded (corresponding to short exposures) 
 \item needed radius of curvature to operate when $r_0=7$~cm: 14~m
 \item estimated Strehl at 500~nm wavelength when $r_0=7$~cm: 0.50
 \item the electrode voltages are well balanced. The voltages on the outermost ring are estimated to be about 70\% of the inner electrodes, which was judged to be needed to compensate for the constraints imposed by the mounting of the DM
\end{itemize}

The results obtained were verified by Marcos van Dam at Flat Wavefronts New Zealand in May 2011, using independently developed software (YAO). In particular, he confirmed that the interaction matrix is very well conditioned \footnote{\url{https://en.wikipedia.org/wiki/Condition_number}} and that there is only one invisible mode: the piston mode. The expected Strehl at 500~nm was predicted to be 0.47 when $r_0=7$~cm at 500~nm, which is close to what was found with our simulation software. This Strehl corresponds to 29\% at 390~nm and 77\% at 850~nm, which are wavelengths of major interest for science observations with CHROMIS and CRISP.

The previous references to seeing conditions corresponding to $r_0=7$~cm deserve a comment. This corresponds roughly to the limit below which the granulation pattern becomes so washed out that correlation tracking for many of the subapertures fails entirely - with the WFS reporting large frame to frame jumps in the measured positions. This causes the AO system to switch off, though locking with a reduced number (low-order) modes may still be possible. We therefore considered it contra productive to require the AO system to have a stroke above that needed for stable lock when $r_0<7$~cm, since larger stroke in practice leads to a lower first resonance frequency (see Sect. \ref{sect:architecture}).

\subsection{Wavefront sensor design}\label{sect:WFS_design}

Figure \ref{fig:Re-imaging} shows the location of the wavefront sensor on the red (CRISP) beam. The beamsplitter cube deflecting light to the WFS is located approximately 30~cm from the focal plane of the SST secondary optical system, which delivers an F/46 beam. This location is intended to be as close as possible to the focal plane of the science instrumentation (CRISP). The WFS consists of an interference filter, a field stop, a collimator lens, two microlens arrays, and a CMOS camera, with the following characteristics:
\begin{description}
\item[\bf Interference filter,] 10~nm full width at half maximum (FWHM), 550~nm center wavelength
\item[\bf Field stop,] equivalent to a field-of-view (FOV) of 18\arcsec$\times$16\arcsec, or 38$\times$34 pixels on the WFS camera (32$\times$32 pixels are used for cross-correlations). At the same location, a pinhole for calibrating the relative subimage positions of the WFS can be folded into the beam
\item[\bf Collimator lens,] Edmund doublet F47-648 with 275~mm (nominal) focal length
\item[\bf Microlens array~1,] from Smart Microoptical Solutions (SMOS) with 38.8~mm focal length. 85 hexagonal subapertures inside pupil, horizontal/vertical pitch 0.544~mm/0.471~mm.
\item[\bf Microlens array~2,] from SMOS. 253 hexagonal subapertures inside pupil, horizontal/vertical pitch 0.326~mm/0.283~mm and focal length 38.8~mm, 
 \item[\bf WFS camera,] Eosens CL MC1362 from Mikrotron GmbH with 14~\textmu{}m pixels. Number of pixels read out: 440$\times$400 out of the 1280$\times$1024 provided by the camera. The maximum frame rate with the number of pixels used is slightly above 2~kHz.
\end{description}
The WFS pupil diameter and the pitch of the microlens arrays are set by by the camera pixel size and the number of pixels needed to cover the (relatively large) FOV of each sub-image. Given the input F/46 beam, the required pupil diameter also defines the focal length of the collimator lens. The focal length of the microlenses is then set by the collimator focal length, the pixel size of the camera and the required image scale.

The design of the Microlens array~1 is such that the microlenses avoid the edge of the pupil, as shown in the upper panel of Fig.~\ref{fig:AO_WFS}. Tests of the collimator lens after delivery revealed that its focal length was 2.3\% shorter than nominal. To compensate for that and a small mismatch between the calculated and measured pupil diameter, the pitch of the microlens arrays was reduced by 5\%. The values above are those of the final design. The image scale is 0\farcs48 per pixel, according to measurements made in May 2016. We note that the quality and contrast of the WFS images appear excellent -- this is likely a consequence of the small number of mirrors and high quality of the SST primary optical system, and the high quality of the microlenses from SMOS.

\subsection{Calibrations} 
\label{sect:calibrations}

\subsubsection{Calibration pinholes}
\label{sect:calibrations}

\begin{figure*}
\centering
\begin{subfigure}{\textwidth}
  \centering
  \includegraphics[angle=0, width=0.99\textwidth,clip]{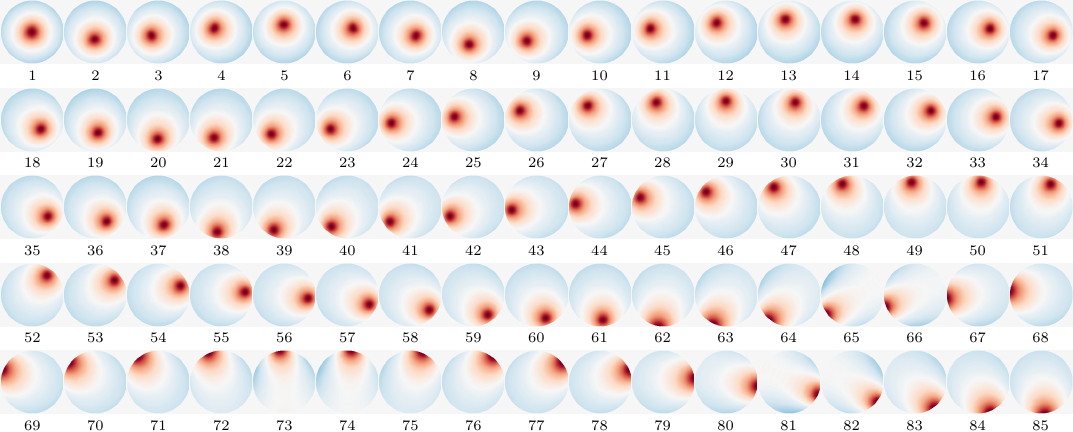}
  \caption{}
\end{subfigure}
\begin{subfigure}{\textwidth}
  \includegraphics[angle=0, width=0.99\textwidth,clip]{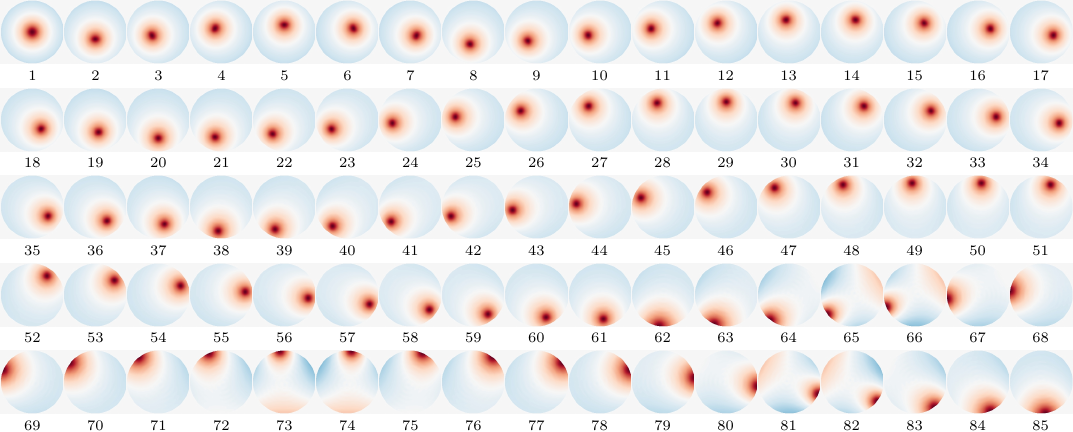}
  \caption{}
\end{subfigure}
\caption{85 electrode influence functions, $E_i(r,\theta)$. The influence functions (see Appendix~\ref{sec:control-modes}), correspond to the matrix $\mathbf{C}_{\text{h}}$ (see text), in (a) as obtained from the calibration of the DM using sunlight on a pinhole located in front of the DM, and monitoring the measured pinhole positions from the 253-lenslet SH WFS, shown in Fig.~\ref{fig:AO_WFS}, while the AO computer outputs a saw-tooth voltage on one electrode at the time. In (b) the same influence functions as calculated theoretically from the electrode layout and the thin plate model (implemented as the solution to the Poisson equation, taking into account the three mounting points of the DM). The color scaling in the upper panel is consistent with that in the lower panel. {\gs Sentence removed!!}
}
\label{fig:DM_response}
\end{figure*}

The AO system uses sunlight for all its calibrations. At the WFS focal plane, there is a pinhole with a diameter equivalent to less than 1\arcsec{} that can be inserted and that is used to define the relative sub-image positions for a perfectly ``flat'' wavefront, i.e., for which the wavefront from the telescope is identical to that produced by the WFS itself. This pinhole is illuminated with (focused) sunlight.

At the primary focal plane, just below the field lens and exit vacuum window and in front of the DM, is the focal plane of the primary optical system of SST. At this location, there is a slide containing various field stops, a grid of pinholes that are used for focusing and alignment (see Sect.~\ref{sect:focusing}), and a pinhole that is used to establish the interaction matrix
of the AO system, as described in the following Section.

\subsubsection{Establishing the control matrix}
\label{sec:establ-contr-matr}

The AO system operates in an optimum fashion through the implementation of control modes. By control modes, we here mean the modes produced by the DM that are orthogonal over the pupil, can be sensed by the SH WFS, and that have amplitudes that are statistically independent (uncorrelated) with turbulence induced seeing. To further optimize the performance of the closed loop performance, we establish the system control modes with the aid of a separate high-resolution wavefront sensor. To do this, we first establish the influence functions from each electrode as follows:

As described in Sect.~\ref{sect:WFS_design}, the FOV of the WFS is very large, about 18\arcsec$\times$16\arcsec{} to allow cross-correlations with a 15\arcsec$\times$15\arcsec{} ``live'' image and a 12\arcsec$\times$12\arcsec{} reference image. This FOV is much larger than needed to establish the interaction matrix of the AO system with a pinhole image. This allows a straightforward implementation of a second more highly resolving WFS by simply replacing the 85-subaperture microlens array with a second array that has 253 lenslets within the 98~cm pupil. The second microlens array has exactly the same focal length as the first microlens array, and its use is implemented simply by mounting both microlens arrays on a motorized x,y stage that is controlled from the AO computer. 

\begin{figure*}
\center
\includegraphics[bb=27 52 566 700,angle=-90, width=0.24\textwidth,clip]{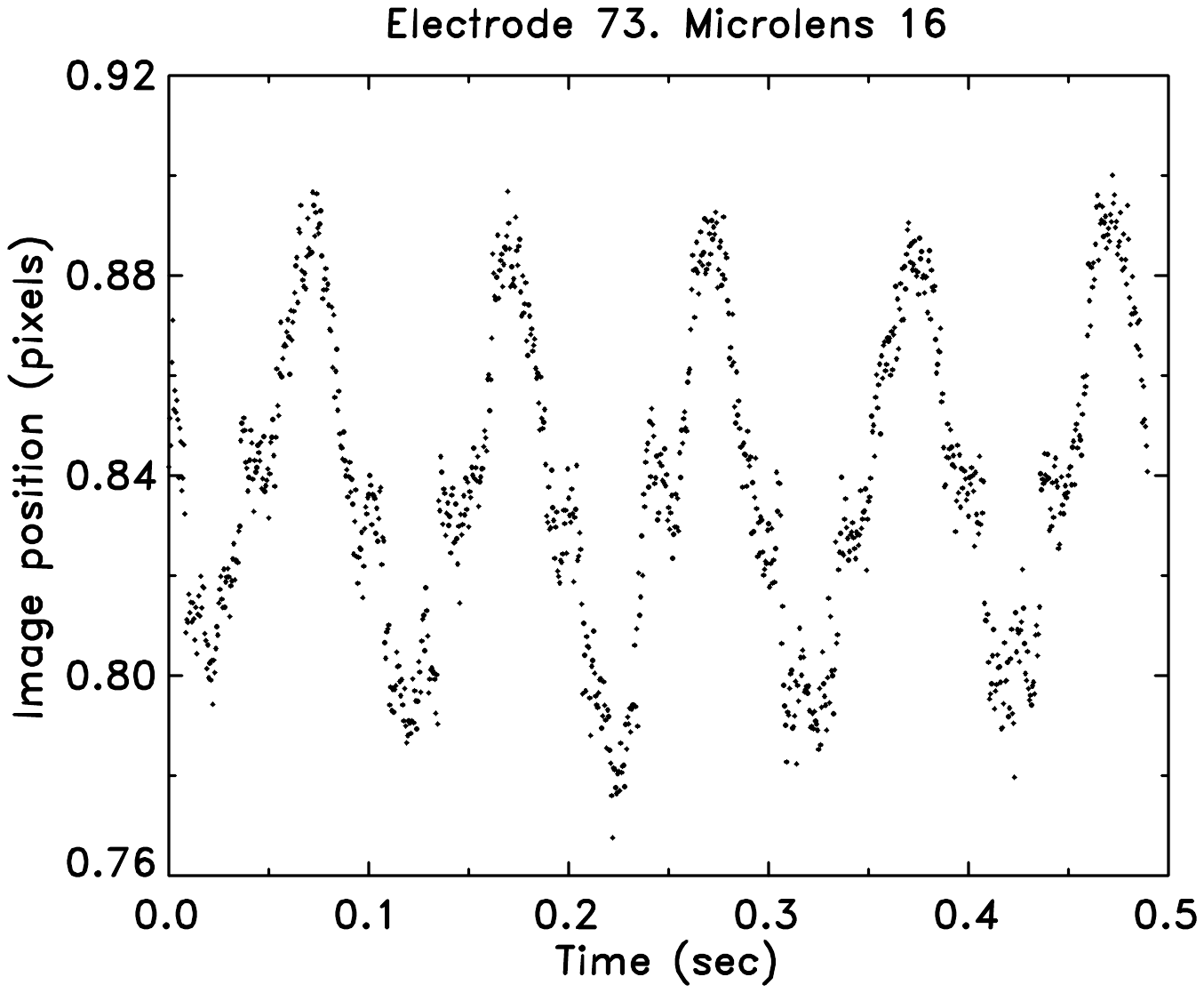}
\includegraphics[bb=27 52 566 700,angle=-90, width=0.24\textwidth,clip]{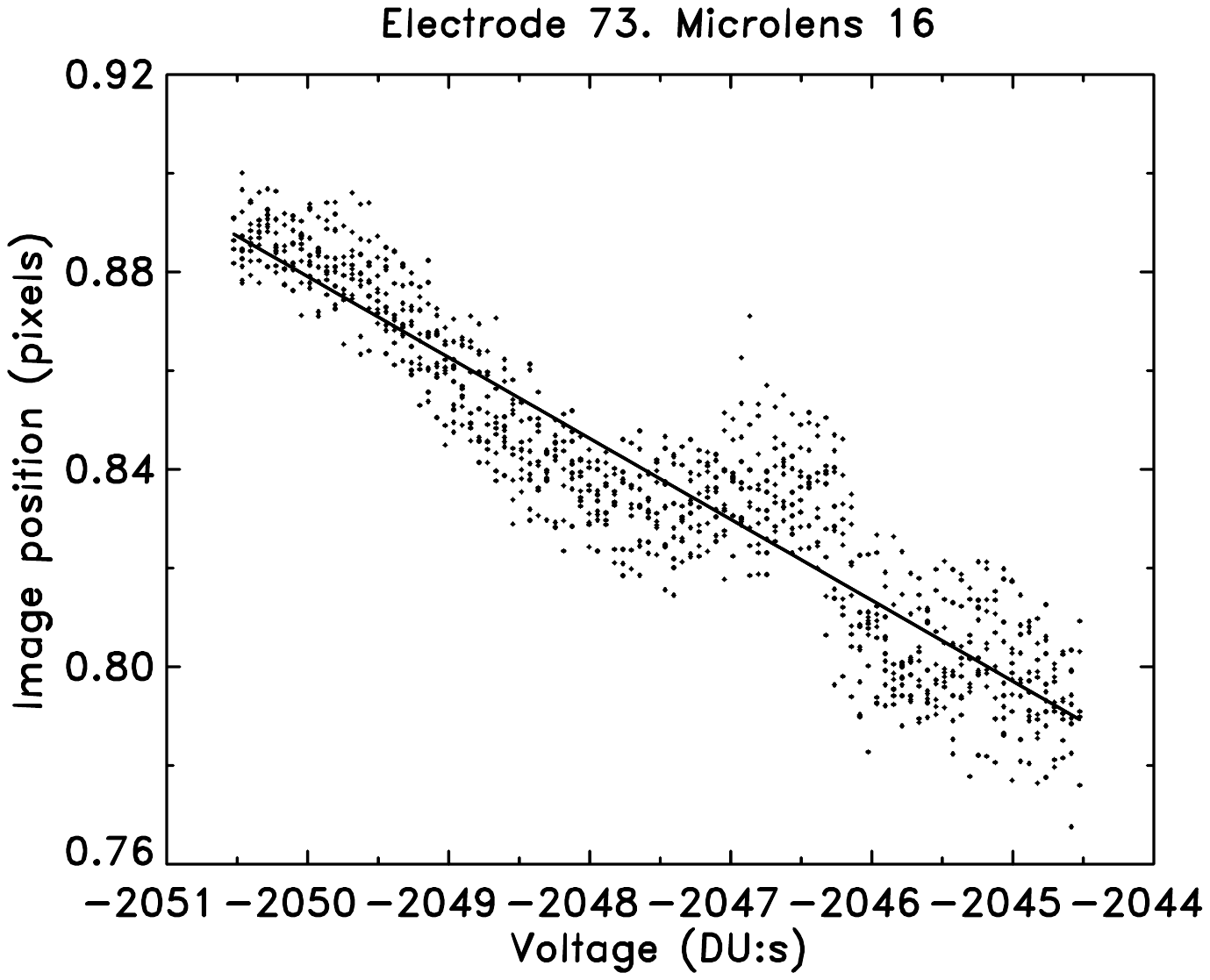}
\includegraphics[bb=27 52 566 700,angle=-90, width=0.24\textwidth,clip]{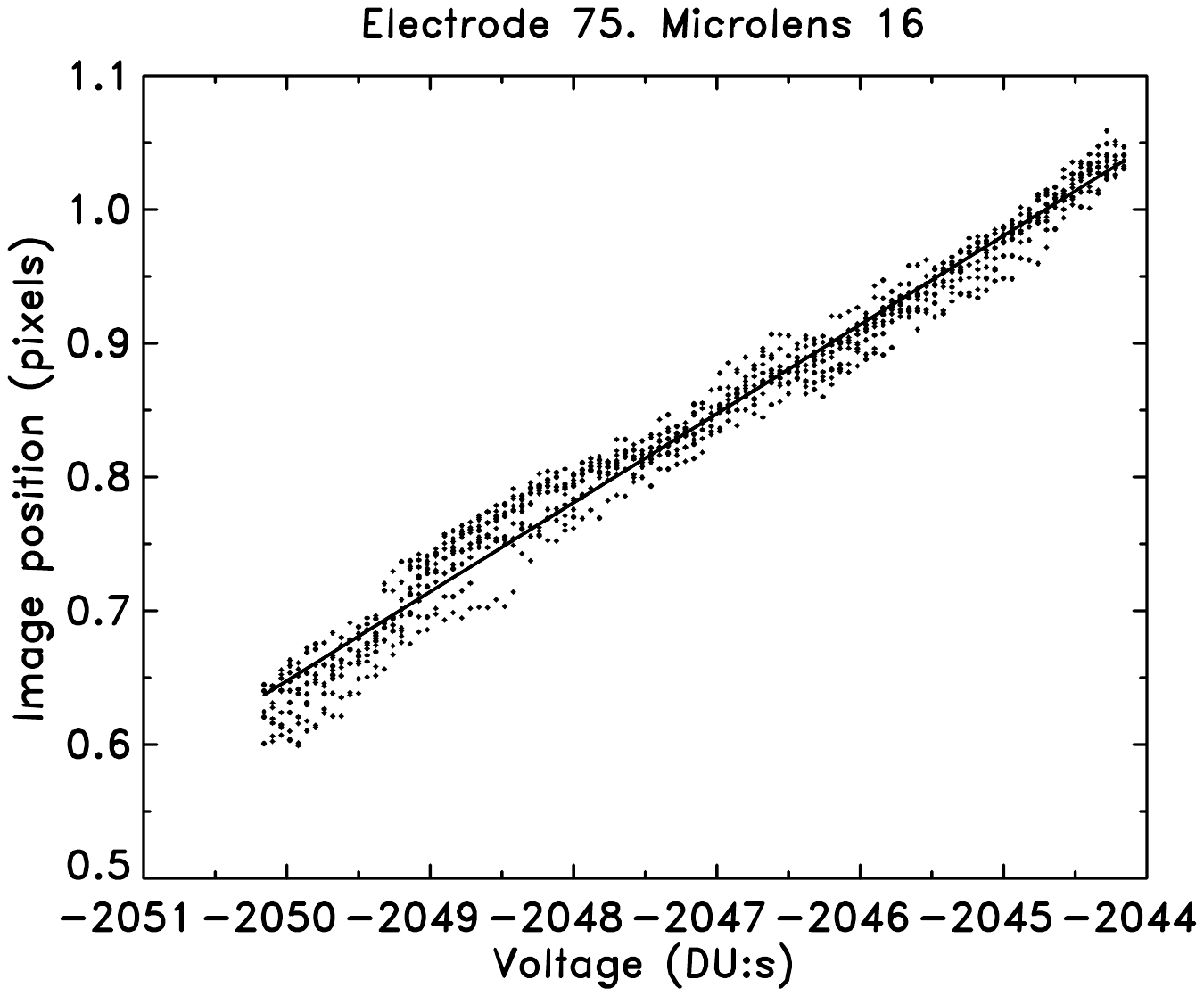}
\includegraphics[bb=27 52 566 700,angle=-90, width=0.24\textwidth,clip]{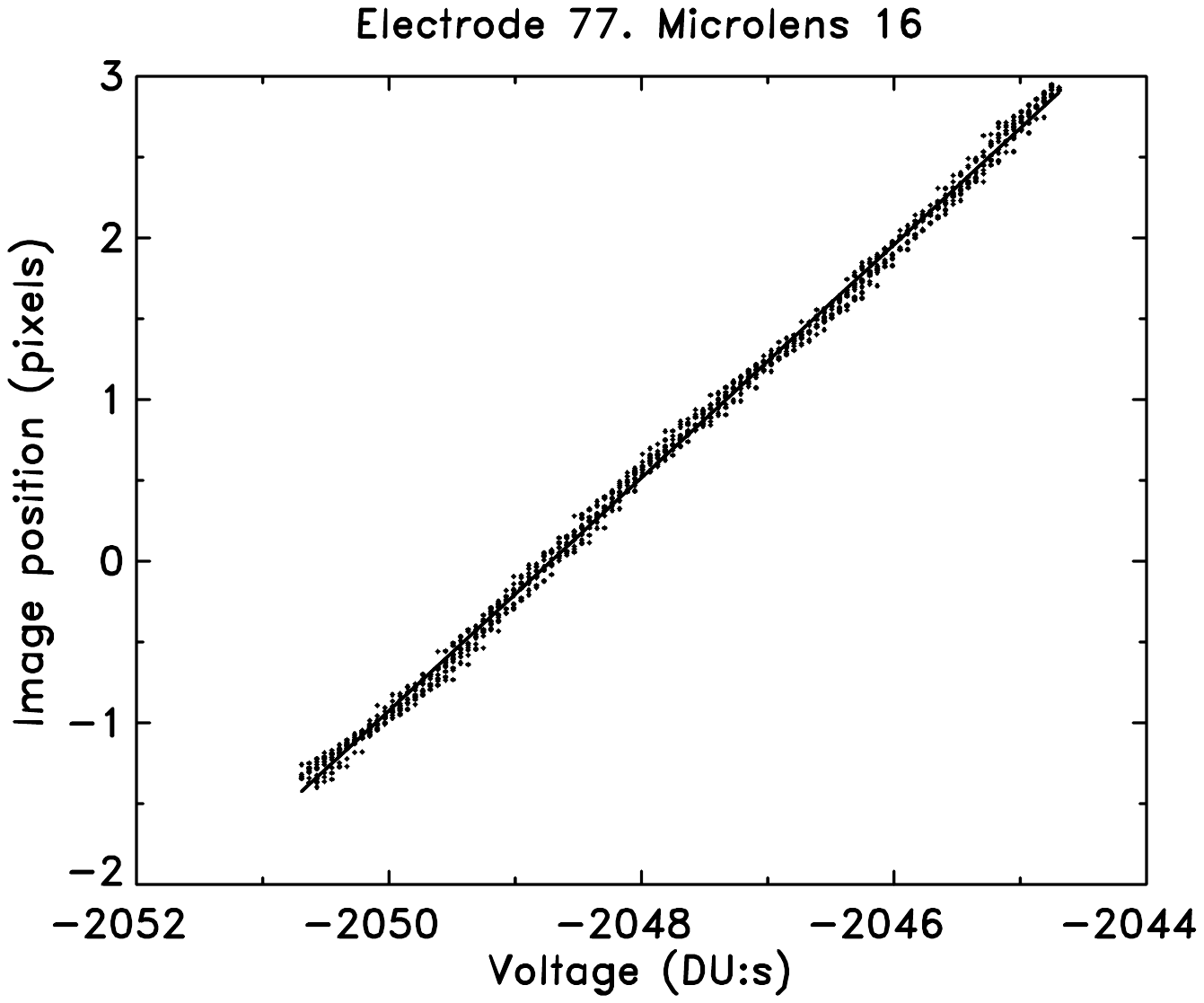}
 \caption{
Measured positions of the pinhole image from subimage 16 (out of 253) as response to the 10~Hz saw-tooth wave of voltages applied to electrodes 73, 75 and 77. Note that the amplitude response to electrode 73 results in a movement of only $\pm 0.06$ pixels or 0\farcs03 , yet the saw-tooth wave of image positions is well above the noise level. The numbering of the electrodes is explained in Fig.~\ref{fig:DM_response} and their relations to microlens 16 are indicated in Fig.~\ref{fig:AO_WFS}.
}
\label{fig:DM_calibration}
\end{figure*}

With the 253-subaperture (high-resolution) WFS, we obtain the interaction matrix by producing a 10~Hz saw-tooth wave of voltages on each electrode separately during a period of 0.5~sec., such that the total time needed to calibrate the 85 electrodes is less than 1~min. In parallel, the software is used to measure the positions of the pinhole images from all 253 subapertures at 2 kHz frame rate. This produces 85 thousand $(x,y)$ positions per subaperture, or a total of 21 million measured positions,  which are post-processed with software that fits straight lines to the measured positions as functions of voltage. The 43010 measured slopes of the fits are stored as elements of the ``high-resolution'' (in terms of number of subapertures) interaction
matrix $\mathbf{C}_{\text{h}}$, the processing of which is discussed below.  Examples of these fits are shown in Fig.~\ref{fig:DM_calibration}. The first of these plots show the variation of the measured x-position for sub-image (microlens) 16 as response to the saw-tooth wave of voltages on electrode 73. Note that even though this results in a displacement of the pinhole image of only $\pm 0.06$ pixel or $\pm 0\farcs03$, the variation of the measured pinhole position is well above the noise level. The correlation between voltage and image position is shown in the second plot, which demonstrates the level of confidence we can have in the obtained linear fit to the data. The next two plots show fits with (much) stronger responses to the applied voltages. Reviewing the data from all 253 subapertures and 85 electrodes, we find a mean RMS deviation of the individually measured positions from the corresponding linear fits of approximately 0.018 pixel, or 0\farcs008. Even though this level of noise should have an inconsequential effect on the quality of the measured interaction matrix, we can discern contributions to the noise in the form of occasional artifacts in the measured data (though not in the plots shown), which could suggest very faint electronic interference or possibly occasional timing problems. 

\begin{figure*}
\center
\includegraphics[bb=27 52 566 700,angle=-90, width=0.24\textwidth,clip]{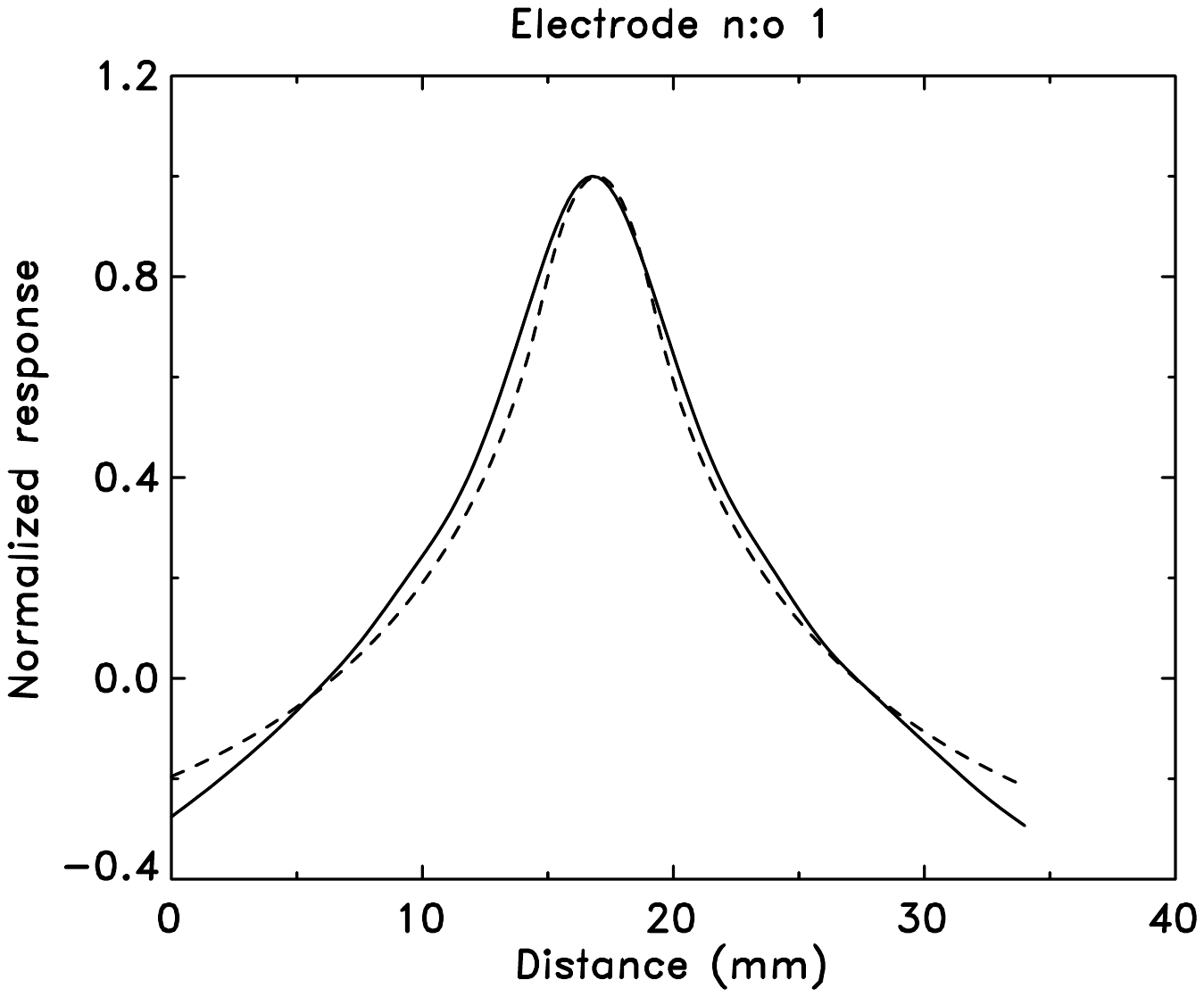}
\includegraphics[bb=27 52 566 700,angle=-90, width=0.24\textwidth,clip]{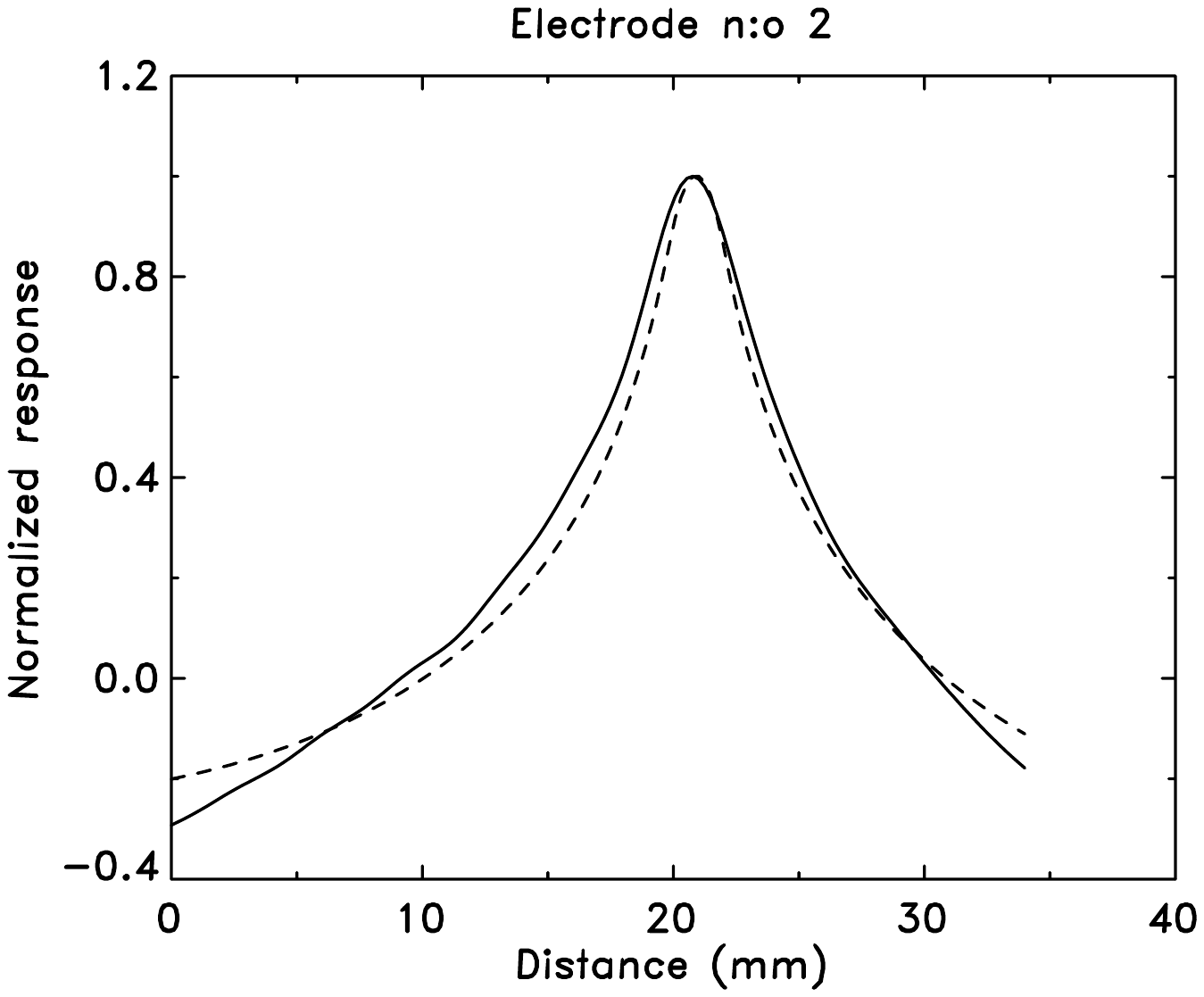}
\includegraphics[bb=27 52 566 700,angle=-90, width=0.24\textwidth,clip]{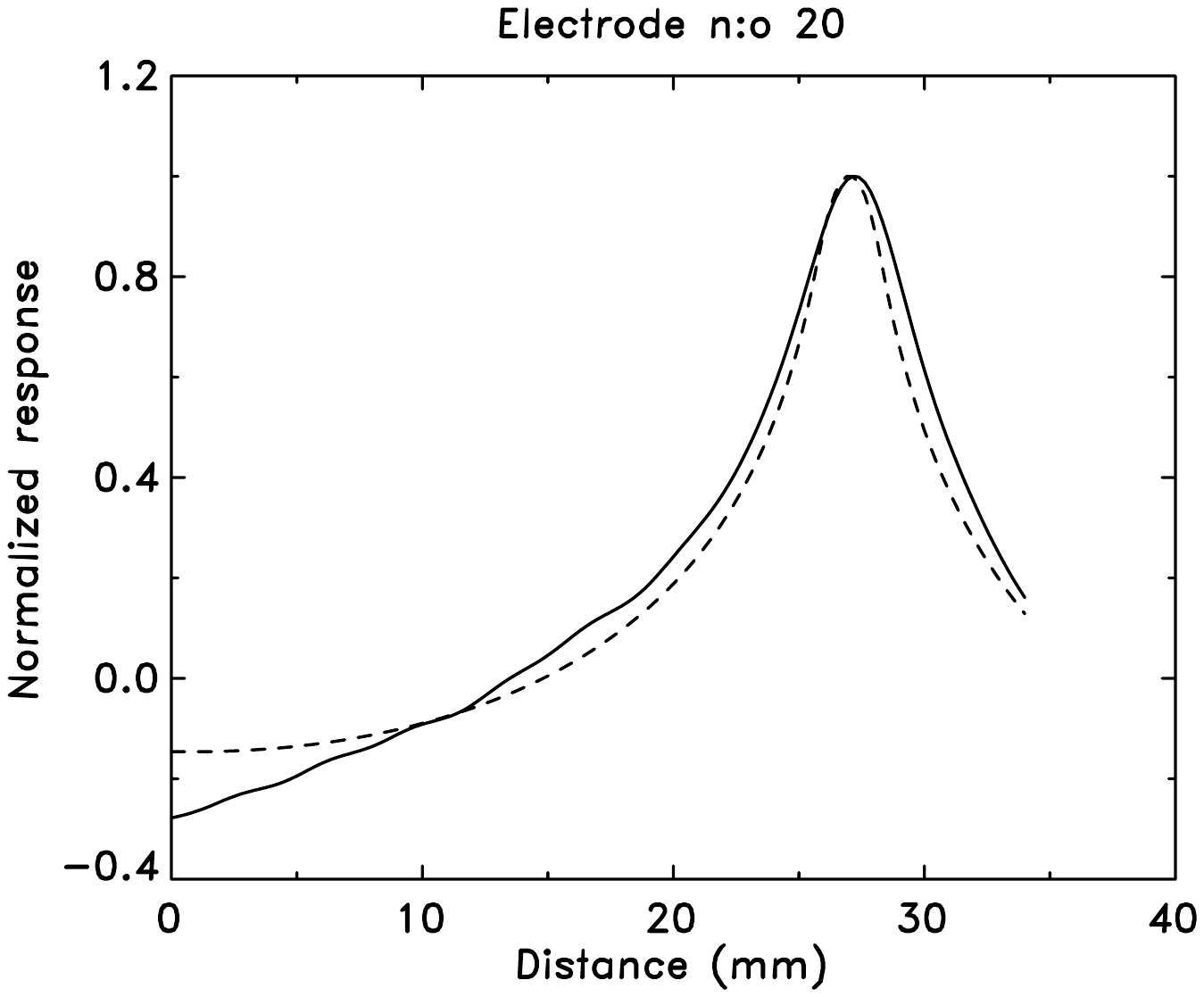}
\includegraphics[bb=27 52 566 700,angle=-90, width=0.24\textwidth,clip]{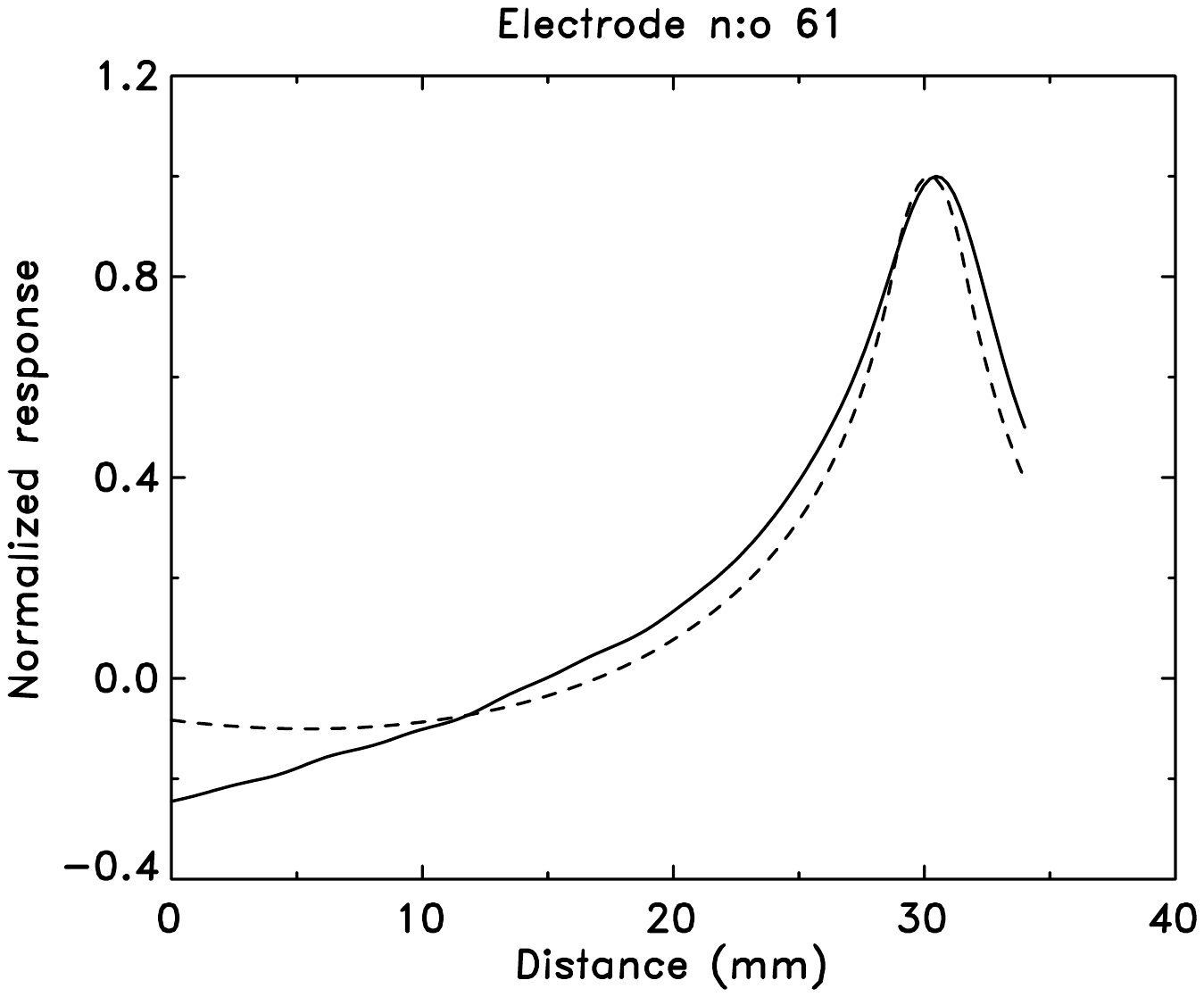}
 \caption{
Vertical cross sections of the response for four of the electrode influence functions shown in Fig.~\ref{fig:DM_response}. The full curve shows the response obtained from the calibration of the DM with the 253-lenslet SH, the dashed curve the response obtained from theoretical simulations. All curves have been normalized to unity at the peak values of the corresponding plot. The numbering of the electrodes is explained in Fig.~\ref{fig:DM_response}.
}
\label{fig:DM_response2}
\end{figure*}

The measured interaction matrix $\mathbf{C}_{\text{h}}$ is used as follows: First, from theoretical expressions for KL functions, we obtain as before a relation between the measured positions $\mathbf{x}$ and the KL coefficients $\mathbf{k}$ the relations:
\begin{equation}
  \mathbf{x}_{\text{h}}=\mathbf{A}_{\text{h}}\cdot \mathbf{k}_{\text{h}},
\end{equation}
and
\begin{equation}
 \mathbf{k}_{\text{h}} = \mathbf{E}_{\text{h}}\cdot \mathbf{x}_{\text{h}},
\end{equation}
where subscript ``h'' refers to the high resolution WFS, and $ \mathbf{E}_{\text{h}}$ is the pseudo inverse of $\mathbf{A}_{\text{h}}$.
Combining this with the interaction matrix $\mathbf{C}_{h}$,
which is related to the vectors of measured positions $\mathbf{x}_{\text{h}}$
and voltages $\mathbf{v}$ (as illustrated in
Fig.~\ref{fig:DM_calibration}) through 
\begin{equation}
 \mathbf{x}_{\text{h}}=\mathbf{C}_{\text{h}}\cdot \mathbf{v},
\end{equation}
we obtain 
\begin{equation}
 \mathbf{k}_{\text{h}} = (\mathbf{E}_{\text{h}}\cdot \mathbf{C}_{\text{h}})\cdot \mathbf{v}
 =\ifmatrix_{\text{h}}\cdot \mathbf{v} .
 \label{eq:3}
\end{equation}
The matrix $\ifmatrix_{\text{h}}$ defines the electrode influence functions, expressed in terms of a (long) vector of KL-coefficients, for each of the 85 electrodes of the DM in the columns of $\ifmatrix_{\text{h}}$. Tests with SVD inversions demonstrate that we can measure 230 KL modes with the 253-subaperture WFS. Summing the KL functions corresponding to the 230 KL coefficients for each electrode,
we obtain the electrode influence functions coded in colour in Fig.~\ref{fig:DM_response}. We note that these influence functions appear clean and noise-free, and that the influence functions from the different electrodes at the same radial distance appear consistent with, and represent simply rotated versions of, each other. To convey further confidence in these empirically determined influence functions, we show in the lower panel the corresponding functions obtained theoretically, using the Poisson equation to solve the thin-plate problem discussed briefly in Sect.~\ref{sect:electrode_layout}. To the eye, the two panels appear almost identical. Figure \ref{fig:DM_response2} shows vertical cuts of the response functions through the brightest pixel in Fig.~\ref{fig:DM_response} for four of the electrodes. Evidently, the measured and calculated responses agree very well close to the electrodes but less so far from the electrode. The agreement nevertheless must be considered quite good and this lends support to the methodology of designing the electrode layout, discussed in Sect.~\ref{sect:electrode_layout}. 

\begin{figure*}
\center
\includegraphics[angle=0, width=0.99\textwidth,clip]{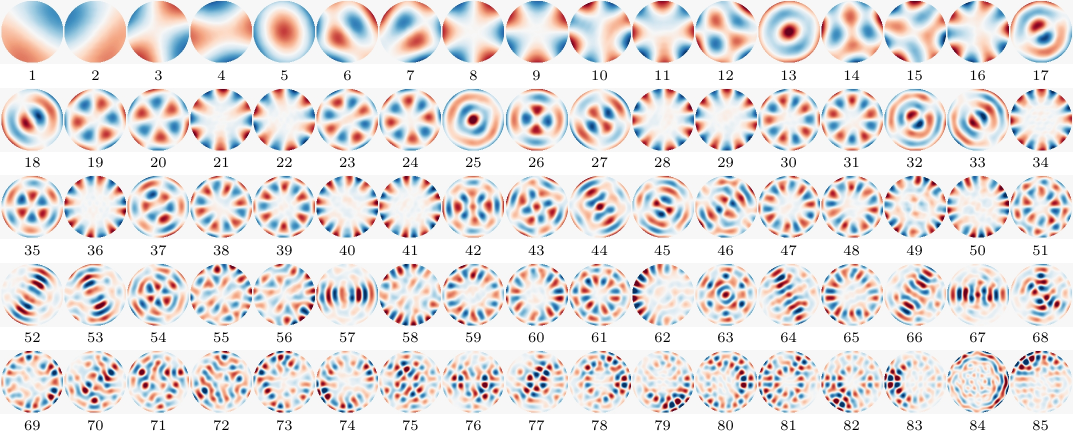}
 \caption{
The 85 control modes $S_i(r,\theta)$, obtained from the electrode influence functions shown in Fig.~\ref{fig:DM_response}, and using the double diagonalization method described by \protect\citet{2000SPIE.4007..620L} - for further details see Appendix~\ref{sec:control-modes}. The modes are ordered according to their expected (descending) variances. The intensity scaling is independent for each mode.
}
\label{fig:DM_control}
\end{figure*}

\begin{figure}
\center
\includegraphics[bb=30 45 710 550,angle=0, width=0.95\linewidth,clip]{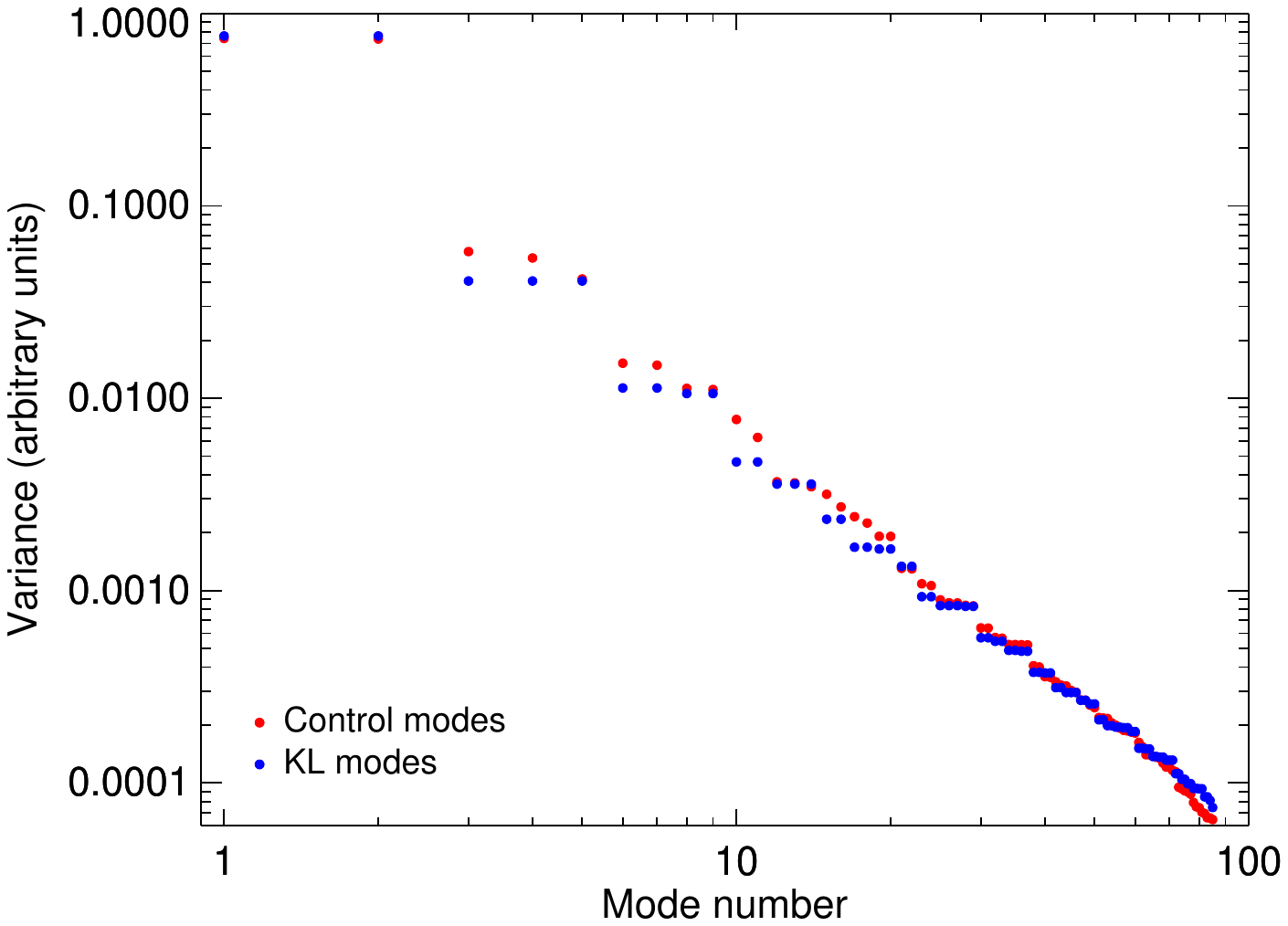}
 \caption{
Expected mode variances for the 85 control modes shown in Fig.~\ref{fig:DM_control} (red dots), compared to the expected variances for pure KL modes (blue dots). 
}
\label{fig:DM_control2}
\end{figure}

To be optimal for AO control, the 85 electrode influence functions shown in Fig.~\ref{fig:DM_response} must be both orthonormalized over the pupil and their amplitudes made statistically independent for turbulent seeing \citep{1991SPIE.1542...34G,1994A&A...291..337G,1999aoa..book...91R}. This is implemented using the double diagonalization method described by \citet{2000SPIE.4007..620L}, see Appendix~\ref{sec:control-modes}.
Using the $\ifmatrix_{\text{h}}$ matrix and well known properties of the KL functions, we obtain from the 253-subaperture WFS the important matrix $\cmmatrix$, where the columns are the vectors of voltages needed to produce the control modes, $S_i(r,\varphi)$, shown in Fig.~\ref{fig:DM_control}. The corresponding atmospheric variances for the control modes are plotted in Fig.~\ref{fig:DM_control2}.

The matrix $\cmmatrix$ connects the vector of mode coefficients $\mathbf{m}$ to unique combinations of DM voltages $\mathbf{v}$ through 
\begin{equation}
\label{eq:v_from_m}
   \mathbf{v} = \cmmatrix \cdot \mathbf{m},
\end{equation}
where $\cmmatrix$ is an 85$\times$85 element matrix. The above 253-subaperture WFS calibration is combined with an 85-subaperture WFS (which is the one used in closed loop) calibration between voltages and $(x,y)$ positions that provides the interaction
matrix $\mathbf{C}$
\begin{equation}
  \mathbf{x} = \mathbf{C}\cdot \mathbf{v},
\end{equation}
Combining the previous two equations gives 
\begin{equation}
 \mathbf{x} = (\mathbf{C}\cdot \cmmatrix) \cdot \mathbf{m},
\end{equation}
which is inverted to enable the calculation of the mode coefficients from the measured $(x,y)$ positions
\begin{equation}
 \mathbf{m} = \mathbf{F} \cdot \mathbf{x},
 \label{eq:4}
\end{equation}
where $\mathbf{F}$ is the pseudo inverse of $(\mathbf{C}\cdot \cmmatrix)$, obtained with SVD methods. This equation is used to calculate the mode coefficients in closed loop and these coefficients and Eq.~\ref{eq:v_from_m} are used to calculate the required DM voltages. As demonstrated above, the servo loop thus is determined by a combination of calibrations using both the ``high-resolution'' 253-subaperture WFS and the 85-subaperture WFS, where the former is used to ``identify'' the appropriate combination of voltages for each of the control modes, and the latter to ``connect'' that identification to the 85-subaperture WFS, which is the one used in closed loop.

{\gs We are still short of a strategy for a proper analysis to demonstrate to what extent the 253-subaperture WFS and the numerical methods described here improves the overall performance of the AO system, as compared an identical system with conventional (85-subaperture) calibration. On the one hand (as pointed out previously), the "smoothness" of wavefronts generated by monomorph and bimorph mirrors implies that the additional high-order modes captured with the 253-subaperture WFS will be small in magnitude. Thus the method can be expected to be of (much) higher importance when used to  characterize a deformable mirror with more localized influence functions, such as piezo-stack mirrors. On the other hand, simply the use of more WFS subapertures to bring down the noise in the measured wavefronts should help to improve the overall performance of the AO system. These aspects deserve further investigation.}  

\subsubsection{Tip-tilt correction}
The first two control modes of the DM are very close to the Zernike tip-tilt modes and it would seem natural to use the DM for that. However, for several reasons we use a separate tip-tilt mirror and separate software and computer. One major reason is to allow observers to change pointing arbitrarily while using the AO system in closed loop, such that they can choose a science target while examining a potential candidate target or updating the pointing at the highest possible spatial resolution. Another reason is that for most scientific programs, it is of paramount importance to retain the pointing even during brief periods of very bad seeing, and such robustness of pointing is best achieved with correlation tracking using a very large FOV.  Furthermore, the science priorities usually dictate the need for stable pointing over a large (1 arc min) FOV, rather than highly performing pointing over a small FOV.  This dictates the use of a correlation tracker with the largest possible field of view to average out the effects of differential image motion (anisoplanatism) over the science field of view. Also, by using the tip-tilt mirror to fold the incoming vertical beam by 60\degr, we can use the DM at an angle of incidence of only 15\degr. Finally, compensating for tip-tilt with the DM will consume some of the stroke that is needed for higher-order aberrations in less favorable seeing conditions. We therefore use a separate camera and correlation tracker computer with processing of 48$\times$48 pixel live images against a 64$\times$64 pixel reference image at 2~kHZ update rate. 


\subsubsection{Focusing science cameras}
\label{sect:focusing}
\begin{figure}
\center
\includegraphics[bb=24 8 493 358,angle=0, width=0.95\linewidth,clip]{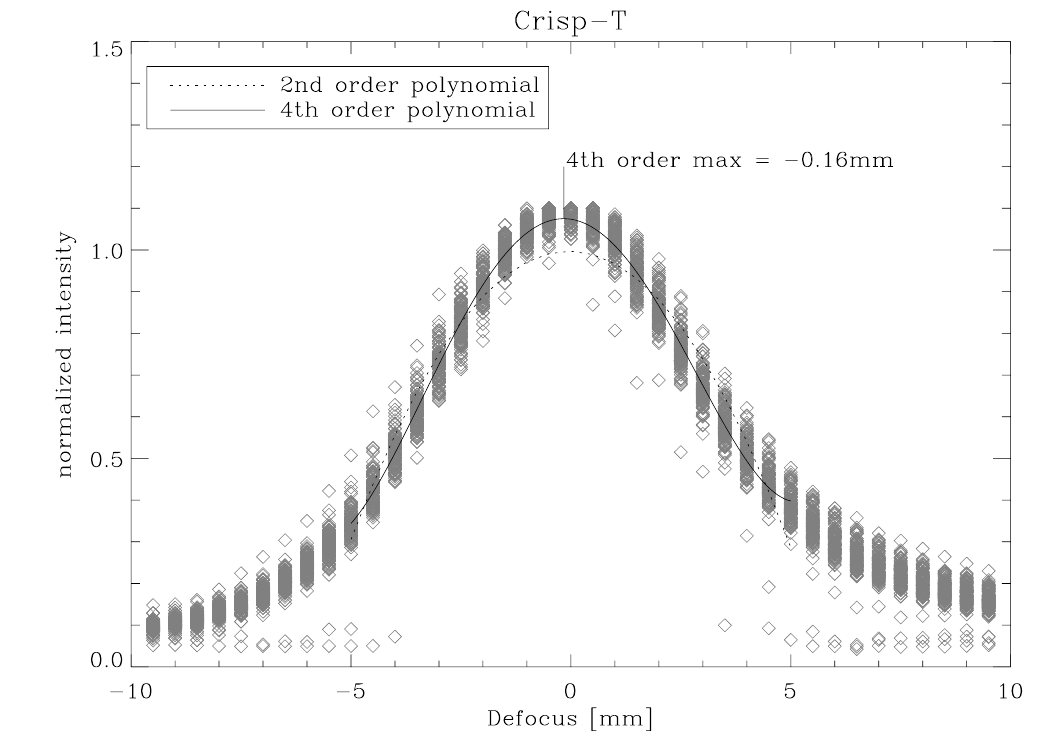}
 \caption{
 Focus plot for one of the CRISP narrowband cameras, containing data from all 121 pinholes within the FOV. A few of the pinholes are partly blocked, resulting in the outliers shown. The remaining data are fitted to a fourth-order polynomial that results in a best-fit focus position of $-0.16$~mm. A second-order polynomial provides a much worse fit to the data but the resulting focus position is almost the same, $-0.04$~mm. It is estimated that the accuracy of the method is better than $\pm$0.3~mm.
}
\label{fig:cam_focus}
\end{figure}

An important advantage of the DM electrode layout and its (monomorph) architecture is the ability to accurately reproduce the focus mode. We use this to implement a precise method for focusing the science cameras. This is done by using the AO system locked on one of the pinholes of the pinhole array located just below the vacuum system (see Sect.~\ref{sect:calibrations}), while stepping through a series of focus shifts during which the software of the science cameras synchronizes measurements of the peak pinhole intensity with the shifting of the focus by the AO system. The multiple pinholes of the array provide focus information over the entire FOV, and not just at its center.  When the intensity measurements are complete, the focus data of the pinhole images are fitted to a parabola and the optimum focus position is found by interpolation. This is done in parallel for all three cameras on the CRISP and CHROMIS beams respectively, using all pinhole images within the FOV. Focusing of the cameras on one beam typically takes about five minutes and involves 2-3 iterations to verify the results. Based on the spread of the best focus position estimated from different pinholes, it is estimated that this procedure leads to a focus error below 0.3~mm in the F/46 beam, corresponding to less than 1/100~wave RMS wavefront error at 630~nm. The procedure provides precise information also about any tilt or curvature of the focal plane, if present, for all science cameras. Figure~\ref{fig:cam_focus} shows a typical data set and fit to the data for one of the CRISP cameras.   

\subsection{Near-limb operation}
To facilitate observations near the limb of the Sun, the field stop of the AO can be moved to lock on a different position within the science FOV of the telescope. This allows, for example, the AO to lock on the left side of the telescope FOV where the disk is visible, while the center and right side of the FOV are outside the disk. Since the light intensity level and the granulation contrast are much lower near the limb, this requires very good seeing conditions. The presence of small-scale bright structures, such as faculae or plage near the limb, can help the AO lock.

When changing the AO lock point, the illumination of the Shack-Hartmann sensor changes, and this requires a different dark- and flat-field. To allow for fast changes of the lock point, 9 pre-determined lock points are available for the observers, arranged in a 3 by 3 grid covering the FOV of the SST. During daily calibrations, dark- and flat-fields are taken for each of the 9 lock points, and the AO software will load the appropriate calibration files when the lock point is changed.

\subsection{Seeing measurements}
\label{sec:seeing_measure}
Whenever the SST is pointed to the Sun, the WFS of the AO system automatically measures and logs the seeing quality, as expressed in terms of  Fried's parameter $r_0$. This is done by measuring the relative image motion between the four WFS lenslets indicated in red in Fig.\ref{fig:AO_WFS}. This constitutes an arrangement similar to the Differential Image Motion Monitor (DIMM), used by ESO for night-time measurements of seeing \citep{1990A&A...227..294S}. For such an arrangement, the variance of image motion along the direction of two lenslets is related to $r_0$ through the following expression, given by \citet{1990A&A...227..294S}
 \begin{equation}
\langle (x(s)-x(0))^2\rangle  = 0.358 \lambda^2 r_0^{-5/3} D^{-1/3} (1 - 0.541 (s/D)^{-1/3}) .
\end{equation}
 In this equation, $D$ is the subaperture diameter, $s$ is the separation between the centers of the subapertures, and $\lambda$ the wavelength. When the SST AO system is in closed loop, the effects of the actions of the DM on the measured differential image motions are compensated for. Measurements are made during overlapping two second intervals, such that one measurement of $r_0$ per second is obtained. 

Observations early in the morning, while the Sun is around 15\degr{} elevation or even less, suggest the clear dominance of (only) two seeing layers at La Palma. One of these layers must be close to the ground (with a strength set by the balance between heating of the ground by sunlight and cooling by wind), and the other at very high altitude, likely at the height of jet streams around the tropopause. This conclusion is based on observations made with a FOV of more than 1~arcmin that shows image degradation that is either uniform over the entire field of view, or manifests itself as geometrical distortions and differential blurring at a scale of a few arcsec. Even when the Sun is at very low elevation, there is almost never evidence for a significant seeing layer at an intermediate height, which would manifest itself in the form av time variable large-scale gradients in image quality across the science field of view.

To capture information about the two aforementioned seeing layers, the AO system is made to monitor seeing at two different angular scales. The first uses cross correlations based on the entire FOV of the WFS, or about 12\arcsec$\times$12\arcsec, and the second only 8$\times$8 pixels, or about 4\arcsec$\times$4\arcsec. It seems clear that the seeing measurements made with a large FOV averages out most or nearly all the high-altitude seeing, because the corresponding $r_0$ measurements often show very large values of around 0.5~m or more early in the morning, when it is obvious that the seeing quality is not that good. At the same time, the seeing measurement made with a small FOV shows much smaller and more reasonable $r_0$ values, as illustrated in Fig.~2 of  
\citet{2019A&A...626A..55S}. This paper also contains details of the validation process used to demonstrate that the $r_0$ measurements serve as an excellent indicator of data quality, and (perhaps surprisingly) a good indicator of the Strehl ratio, which is used commonly as a measure of optical quality (for further comments, see Sect.~\ref{sect:Comments}). 

Based on the above discussion, and the observation that image quality almost always changes uniformly over the FOV when the AO system operates in closed loop, we also conclude that the 12\arcsec$\times$12\arcsec{} FOV of the AO WFS is sufficiently large for the AO system to effectively work as a ground-layer AO (GLAO) system. This is important, because excellent compensation also for high-altitude seeing, accomplished with a WFS that operates with a small FOV, would come at the prize of image degradation by high-altitude seeing over almost the entire science FOV!

The $r_0$ measurements logged together with all science data will be used to further improve image reconstruction by providing a statistical compensation for the degradation of science data by high-order KL seeing modes that cannot be compensated for with the AO system, or compensated for individually in post processing \citep{2010A&A...521A..68S,2022A&A...668A.129L}. This will further enhance the quality and fidelity of SST science data. 

\section{AO computer hardware and software}
\label{sect:AO_ware}
\subsection{Hardware}

The camera used for the wavefront sensor is a Mikrotron EoSens MC1362, which has 1280 by 1024 pixels, a pixel size of 14 \textmu{}m, and a global shutter. Using a CameraLink Full interface, the entire area of the sensor can be read out at 500 Hz. By reading out only a region of 440$\times$400 pixels, images can be captured at up to 3723 Hz. The internal image processing functions of the camera are all disabled, except for the black level offset, which is set such that when light to the sensor is blocked, the minimum intensity of an exposed frame is above zero. This is necessary to ensure that proper dark field correction can be performed.

The frame grabber used is an Engineering Design Team PCIe8 DV-C-Link, without any on-board memory installed.
Therefore, incoming data from the camera is written to the RAM of the computer with a negligible delay.

The computer is comprised of an ASUS Sabertooth X58 motherboard with an Intel Core i7 X990 CPU installed, along with 3 GB of DDR3 memory. The CPU has 6 cores, each running at 3.47 GHz. While the CPU supports hyperthreading, it has been disabled, since testing the performance of the system demonstrated that enabling hyperthreading did not increase the throughput, but instead increased the latency of the image processing algorithms. Furthermore, the BIOS has been configured to disable all unused hardware functions of the motherboard, such as the on-board sound card and some USB functionality, as these would generate System Management Interrupts (SMI) that could interrupt a core for up to 1 ms, which would result in unacceptably high latencies for the AO servo loop.

To control the deformable mirror, a PXI crate is connected to the computer via a PCIe4 cable. The PXI crate houses 3 National Instruments PXI-6723 analog output modules. Each module has 32 outputs with a resolution of 13 bits, with a maximum update rate of 800 kS/s. The voltage range of each output is $\pm 10$~V.

The output signals from the three PXI-6723 cards are amplified by two CILAS ED64A  high-voltage amplifiers. These have a fixed gain of 40:1, resulting in a final output to the deformable mirror of up to $\pm 400$~V, and a bandwidth of 5~kHz.

\subsection{Operating system}

The computer runs Debian GNU/Linux, with a custom-built 2.6.32 kernel, with preemption enabled to reduce the impact of interrupt handlers. While the CPU can run in 64-bit mode, the kernel and the OS are 32-bit.
The reason for running a rather outdated 32-bit kernel is that Linux drivers from National Instruments for the PXI-6723 cards do not support more recent Linux versions, nor 64-bit.

\subsection{Minimizing latency}

The latency of the servo loop, which is the time from exposure until the actuation of the deformable mirror, should be as low as possible for optimal performance of the system. Low latency is achieved by the following techniques:

\begin{itemize}
\item As many undesirable sources as possible of hardware and software management interrupts have been disabled.
\item The kernel has been compiled with preemption enabled.
\item The software threads implementing the servo loop are running with real-time priority, so they are not interrupted by other, lower priority processes.
\item Thread-affinity is used to bind each thread to its own CPU core, to ensure the per-core L1 and L2 caches of the CPU are not needlessly invalidated by threads migrating to different cores.
\item The frame grabber has no memory of its own, and writes the data received from the camera directly into the RAM of the computer.
\item For cross-correlations, a reference image is selected from the previous exposure (recorded approximately 0.5~ms earlier), which allows processing to start as soon as the first subimage of the present exposure is in memory (see below). Since this reference image is recorded with a time difference of only 0.5~ms relative to the presently exposed subimages, the AO system can remain in closed loop even when the observer changes the telescope pointing, or if correlation tracking is switched off.
\item Processing of a subimage starts as soon as the last pixel of that particular subimage has been written to RAM. This allows read-out from the camera and image processing of the subimages to proceed in parallel, such that the computational delay from the image processing is only 72\textmu{}s after reading out the last row of pixels from the camera.
\end{itemize}

The first four items ensure that each thread can run at full speed with minimal interruptions. The last two items ensure that by the time that the last row of subimages has been read out from the camera, most of the image processing has already been completed. Once the remaining row is processed, the calculated wavefront deformation is converted to a set of voltages to be applied to the deformable mirror.

Each subimage is processed on a single core in approximately 17~\textmu{}s. However, 5 CPU cores are used for image processing
and each core is assigned 17 of the 85 subimages.

\begin{figure}
\includegraphics[angle=0, width=\linewidth,clip]{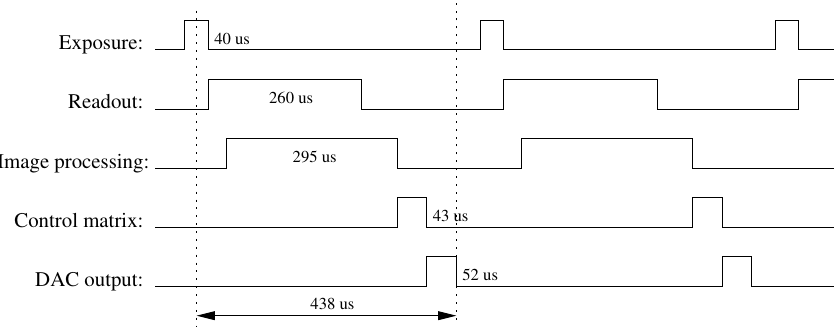}
 \caption{
Timing diagram of the processes involving the WFS exposure and readout, and processing of the WFS data.
}
\label{fig:ao_timeline}
\end{figure}

A graphic representation of the timing of the camera exposure and readout, and the processing, is shown in Fig.~\ref{fig:ao_timeline}. The following contributions to latency have been identified and quantified:
\begin{itemize}
\item One half of the exposure time ($0.5 \cdot 40~\text{µs} = 20$~\textmu{}s).  
\item Read out of all subimages (251~\textmu{}s).
\item Image processing after reading out the last subimage (72~\textmu{}s).
\item Control matrix multiplications (43~\textmu{}s).
\item Transfer of desired voltages to DAC cards (52~\textmu{}s).
\end{itemize}

Total latency: 20+251+72+43+52 = 438~\textmu{}s. Added to this is any latency contributed by the ED64A high-voltage amplifier and the DM itself. 


\subsection{Using multimedia instructions for cross-correlations}

The performance of the AO has been possible in a large part due to the use of MMX instructions.
In particular, the Intel MMX instruction set provides the instruction PSADBW \citep[which is similar to the PERR instruction used in the AO system for SVST, see][]{1999ASPC..183..231S,2000SPIE.4007..239S} that computes the sum of the absolute difference of the components of two vectors of 8 values, as in the following equation:

\begin{equation}
s = \sum_{1 \leq i \leq 8} | x_i - y_i |
\end{equation}

The values must be unsigned 8-bit integers. This is ideal for our AO, where the subimages are 32 by 32 pixels, the reference image is 24 by 24 pixels, and the output of the camera is 8 bits per pixel. 

To perform the cross-correlations, the ``live'' subimage is shifted by $\pm$4~pixels in steps of 1~pixel relative to the reference image, while accumulating the results of all PSADBW calls in an MMX register, resulting in a 9$\times$9 pixel correlation matrix. For each possible image shift, the PSADBW instructions is called $24\cdot24/8 = 72$ times. The minimum of this correlation matrix is found and the 3$\times$3 pixel matrix around that minimum is extracted, and then used to find subpixel shifts by interpolation based on polynomials of degree 2. However, this correlation matrix has a triangular appearance around its minimum, which delivers unacceptably inaccurate subpixel shifts with 1D or 2D interpolation. In 1993, it was recognized by one of the authors (Scharmer) that this fundamental limitation can be overcome by taking the square of the 3$\times$3 matrix before doing the interpolation. This simple modification of the absolute difference algorithm, which is referred to as the ADF$^2$ algorithm, leads to a dramatic improvement of the subpixel accuracy at a negligible computational cost \citep[for details, see][]{2010A&A...524A..90L}.  


The code is written in C++, and uses compiler intrinsics to generate the PSADBW instruction. The advantage of this compared to hand-written assembly code is that the compiler is free to assign registers and reorder the instructions as it sees fit.

One problem encountered while writing the code is that on AMD processors, which also support the MMX instruction set, unaligned loads from memory are handled correctly and relatively fast, while on the Intel CPU that is used for the AO computer, unaligned loads either return incorrect data, or special instructions for unaligned loads have to be used which turned out to be very slow. Instead, the code generates pre-shifted copies of the reference image once per frame, such that the cross-correlation algorithm can be executed using only aligned loads

Timing tests demonstrate that the 85 cross-correlations take a total of 59~\textmu{}s per frame with the PSADBW instruction, and that an implementation in C++ would take 537~\textmu{}s per frame, such that the PSADBW version is 9.1 times faster than an implementation in pure C++. These tests also demonstrate that dark- and flat-field correction of the images take a non-negligible amount of time, and that the latency could be reduced by up to 70~\textmu{}s with a more efficient implementation.


\subsection{Global and subimage intensity correction}

The amount of light received by the camera varies throughout the day, and also depends on the telescope pointing.
The AO is not very sensitive to global changes in intensity, except when the image is severely under- or over-exposed.
However, since exposure time changes cause slight delays in the camera readout,
this results in brief distortions of the deformable mirror when running in closed-loop.
To ensure that this does not interfere with the recording of science images,
changing the exposure time is not automatic, but instead has to be made by the observer.
A warning is displayed on the screen if the AO detects that the exposure time is not optimal.

Whereas the AO image is normally calibrated using a dark-field and flat-field image,
the intensity levels of individual subimages can vary with respect to each other during the day
due to changing illumination of the Shack-Hartmann lenslets (especially those close to the edge of the pupil).
Since the sum-of-absolute-difference function requires the subimages to all have a similar intensity level,
any differences in illumination are compensated for.
To do this, a running average of the average intensity level of each subimage is calculated.
Once every ten seconds, the flat-field image is updated to ensure equal intensity in all subimages.

\subsection{Stability}
\label{sec:stability}

The AO software has been designed to run unsupervised for extended periods in closed-loop mode, even in extremely variable seeing conditions. This is achieved using the following methods:

\begin{itemize}
\item The AO does not compensate for image motion (tip-tilt), and therefore can run in closed-loop mode even when the observer changes the pointing of the telescope.
\item For every frame, the AO determines the number of subimages with a good lock\footnote{A good lock corresponds to the cross-correlation function returning a realistic image shift. The measured shift is considered realistic if both the minimum value of the 9$\times$9 pixel correlation matrix is in one of the inner 7$\times$7~pixels (and not on a border pixel),
and the result of the sub-pixel interpolation also falls within $\pm 3.5$~pixels.}.   
      If less than 75 subimages have a lock, the AO goes into a temporary open-loop mode, changing the shape of the mirror gradually from the last good shape to a flat shape, as given by the offset voltages (see below). 
      Once 80 or more subimages have a lock, the AO goes back to the normal closed-loop mode.
\item The highest 10 control modes require large voltages when the seeing is bad.
      If the control matrix requires voltages that exceed the maximum limit,
      then the highest order modes are effectively disabled by reducing their gains to very low levels, until the required voltages drop to acceptable levels again.
      Also, if the highest order modes have already caused the servo loop to accumulate a large contribution to the electrode voltages,
      these voltages are slowly removed to increase the voltage ``budget'' for the lower order modes.
\item When in closed loop, the AO system continuously updates the offset voltages, which are the optimum voltages for regaining lock after a period of bad seeing. This is needed to compensate for thermal drifts of the shape of the DM. 
\item The AO can be remotely controlled.
      This can be used by the telescope control software to automatically start and stop the AO when needed, such as when the telescope goes into flat-field mode. Similarly, the CT servo loop is automatically opened by the telescope control software, if the observer changes the telescope pointing. 
\end{itemize}

\subsection{Graphical user interface}

The AO software is split into a background process that implements the actual image processing and servo loop,
and a graphical user interface (GUI) that allows the observer to see the current state of the AO, change its control mode,
and perform simple calibrations. This GUI runs on a separate CPU core, such that display and interaction with the observer
does not interfere with the performance of the AO system.

The GUI shows the live image from the wavefront sensor camera, with the positions of the subimages and the measured image shifts overlaid. It also shows a small representation of the deformable mirror, using various colors to indicate the voltage of each electrode, and a bar graph of the residual amplitudes of the control modes. Various warnings are shown in blinking text, such as when the camera is not properly exposed, or when the AO is not in a normal operating mode. All this makes it straightforward for the observer to continuously monitor the performance of the AO system. 

\section{Comments on performance}
\label{sect:Comments}
\subsection{Image quality}
The performance of a night-time AO system can be evaluated by measuring the peak intensity of a star with the AO system, while alternating between open and closed AO servo loop, which provides a direct measure of the achieved Strehl ratio \citep[for example,][]{1998PASP..110..837R}. With a solar AO system, such simple measurements of the Strehl ratio are not possible. \citet{2007PhDT........41M} and \citet{Marino:10}, see also \citep{2011LRSP....8....2R}, modified the WFS of the AO system of the R.B. Dunn Solar Telescope to lock on bright stars such as Sirius and could confirm the predicted PSF. However, solar AO systems use extended WFS targets (solar granulation or other solar fine structure), with typical FOV of 10\arcsec$\times$10\arcsec{} or more, and virtually all solar fine structure is observed against a luminous background, which makes performance estimates much more difficult.

We have recently demonstrated that measurements of the granulation contrast with a meter-class solar telescope compensating for many (for SST, 50 or more) low-order aberrations with AO (and/or image restoration techniques) can be used as an excellent proxy for Strehl measurements \citep{2019A&A...626A..55S}. This is because an AO system that compensates for many (but yet a limited number) of low-order aberrations, produces a PSF with a well-known halo  \citep{1992ESOC...42..475C,1998aoat.book.....H,2006PASP..118..885B}, which typically has a FWHM that is an order of magnitude larger than that of the diffraction limited PSF of the same telescope. Since solar granulation has a typical scale of about 1\farcs5, it is the wings of the PSF (the "halo") that reduce the granulation contrast, whereas the PSF core leaves the contrast intact. Thus it is the fractional area of the PSF contained in the wings that determines the granulation contrast, and this fraction is approximately $1-S$, where $S$ is the Strehl \citep{2019A&A...626A..55S}. 

Based on the above, we have used measurements of granulation contrast to evaluate the overall image quality achieved with the combination of the SST primary and secondary optical systems, the AO system, and the optics of science instrumentation \citep{2019A&A...626A..55S}. Based on these measurements, we conclude that SST with its present 85-electrode AO system delivers granulation contrasts that in excellent seeing conditions are higher than with any other solar telescope (except possibly DKIST, for which we have no data), consistent with the excellent quality of SST science data demonstrated in numerous publications. However, the origin of an additional 48~nm RMS wavefront error, needed to explain the measured granulation contrast, so far remains unexplained.
  

\begin{figure*}
\center
\includegraphics[angle=0, width=0.9\textwidth,clip]{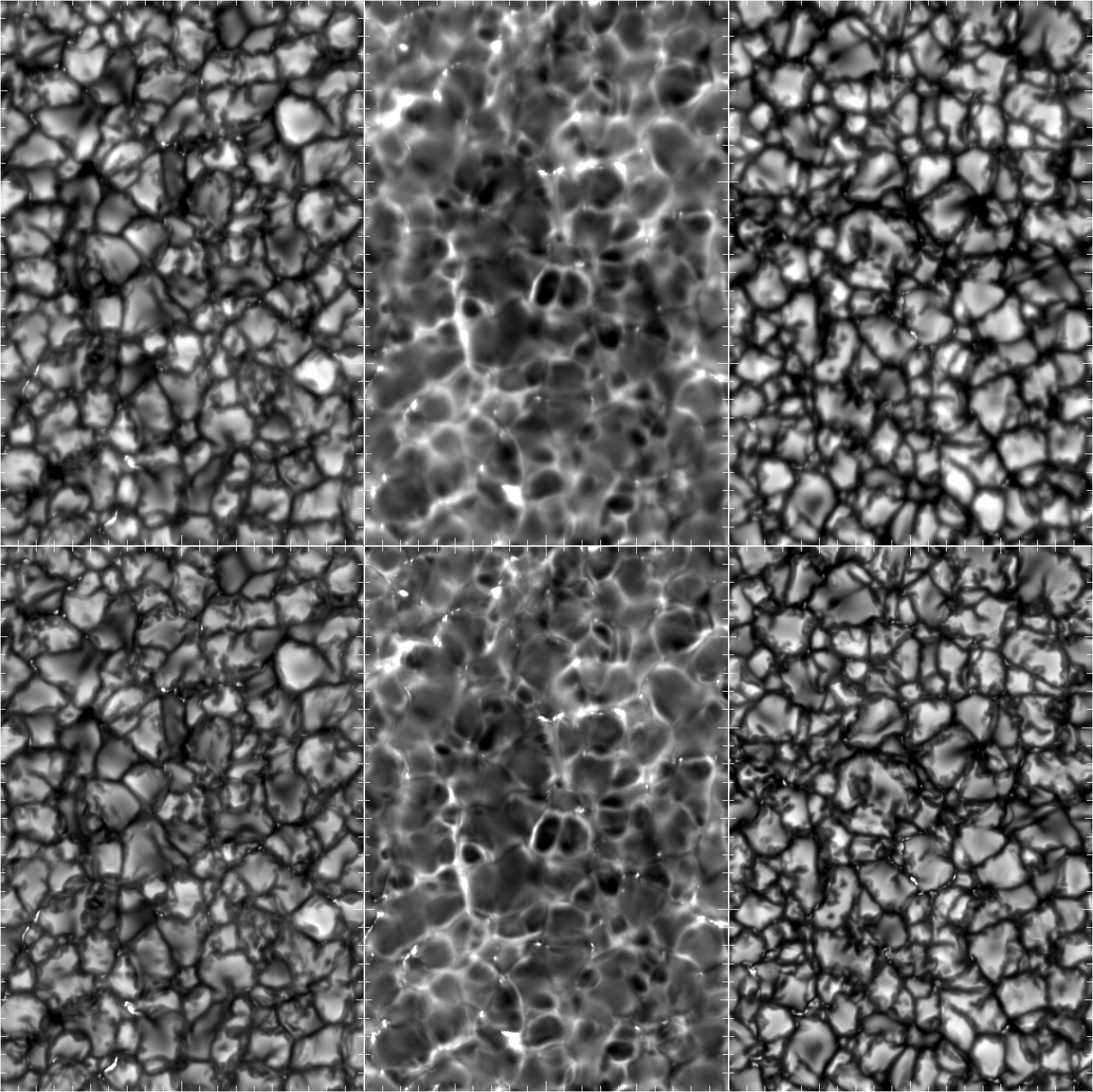}\\[2mm]
 \caption{
   Illustration of the performance of the AO system and the resulting image quality at wavelengths below 400~nm. Top 3 panels show the resulting CHROMIS WB images  after adding up the 100 observed images during 2~seconds and only compensating for relative image motion (tip and tilt) between the individual exposures. The bottom 3 panels show
the same observed images but after applying MFBD image reconstruction to compensate for the 100 most significant KL modes. The wavelengths are (left to right) 395.1~nm (FWHM~1.0 nm), 396.85~nm (Ca II H core, FWHM~0.35 nm), and 399.96~nm (FWHM 0.35~nm). The FOV is 20\arcsec$\times$30\arcsec. The Fried parameter $r_0$ for these data, as measured at 500~nm by the AO system, was (left to right) 30.9~cm, 20.1~cm, and 17.9~cm. At 400~nm, this corresponds to $r_0$ values of 23.6~cm, 15.4~cm, and 13.7~cm, resp. Note the tiny bright points that are indicative of the spatial resolution. 
}
\label{fig:CHROMIS1}
\end{figure*}

As an illustration of the image quality delivered by SST and its 85-electrode AO system in favorable seeing conditions, we show a mosaic of six images in Fig.~\ref{fig:CHROMIS1}, observed at wavelengths at or below 400~nm. The top three images correspond to averages of 100 images recorded during 2~seconds and that are compensated only for image motion (tip-tilt) between the exposures. The three corresponding images at the bottom have been processed with the Multi-Frame Blind Deconvolution (MFBD) method proposed by \citet{2002SPIE.4792..146L} and implemented by \citet{2005SoPh..228..191V}. 

The SST in its present incarnation is arguably the ground-based solar telescope that until now has delivered the highest image quality, in particular at short wavelengths. There must be several factors contributing to this. One possibility is that the site of SST on La Palma is the best known site for solar telescopes. Another possibility is that an optical system like that of  SST, with a small number of mirrors and lenses of excellent optical quality, is a prerequisite for achieving high-fidelity image quality. Expanding on this argument, we conjecture that a telescope with a large number of optical elements (in particular mirrors), each of a quality that is not outstanding, may end up with an unacceptably large variance of accumulated aberrations at scales smaller than what the AO system can compensate for. This may then explain why some other solar telescopes appear to deliver a Strehl (granulation contrast) that is well below that of SST. 

A third factor explaining the high image quality of SST, is that there are (possibly unidentified) aspects of the SST AO system that are particularly important. We can identify the following characteristics that may be of particular relevance:
\begin{itemize}
 \item The documented high optical quality of the DM (Sect.~\ref{sect:architecture})
  \item The high quality and contrast of the WFS images and its large field of view, used for cross correlations (Sect.~\ref{sect:WFS_design}), and effectively turning the AO system into a ground-layer AO (GLAO) system (Sect.~\ref{sec:seeing_measure})
 \item The likely very accurate calibration matrices (Sect.~\ref{sect:calibrations}) and modal control (Appendix~\ref{sec:control-modes})
 \item The location of the beam splitter cube for the AO WFS close to the science focal plane 
 \item The accurate focusing procedures for science cameras (Sect.~\ref{sect:focusing})
 \item The overall stability and robustness of the AO software (Sect.~\ref{sect:AO_ware})
  \item The use of a monomorph DM.
\end{itemize}

We cannot state which of the above explanations and AO characteristics are predominant - to some extent all aspects must play a role - but it is a question of obvious importance for the development of future telescopes, such as EST \citep{2013MmSAI..84..379C,2016SPIE.9908E..09M,2019AdSpR..63.1389J, 2022A&A...666A..21Q}. Evidently, more needs to be done in order to investigate and explain performance limitations of existing solar telescopes, such that we can develop a new generation of telescopes that operate closer to their theoretical limits.

\subsection{Computational aspects and bandwidth improvement}
There are several performance aspects that likely deserve investigation and improvements. One aspect that has not been well investigated is the 0~dB bandwidth of the error transfer function. Tests were made with an early version of the AO system, which used the present 85-lenslet WFS (including all the heavy cross correlations), and the same HV amplifiers as now, but with the previous 37-electrode DM. These tests gave a 0~dB error rejection bandwidth of 130~Hz, but the tests should be repeated with the goal of improving the bandwidth even further. Noting that the AO computer and WFS camera are eight years old, suggests a significant potential for improvement. Comparing the performance of the present AO computer with that of a present-day work station in a similar price range, indicates that the overall latency can be improved by about 50~\textmu{}s at modest cost and effort.  A more ambitious upgrade could involve using a larger and faster WFS camera with a more powerful AO computer that would reduce latency from both readout and image processing. Alternatively, a faster AO computer and camera can be used for cross-correlations with larger FOV to give even more stable performance in poor seeing, and to average out even more of the high-altitude seeing. Other performance aspects that certainly deserve attention is to replace the PXI-6723 cards with DAC cards that support 64-bit Linux -- probably this will lead noticeable speedups as well.

\subsection{Development costs}
We note that the overall architecture of the AO system, hardware and software leads to a system that is cost efficient in terms of development and maintenance. The software of the AO system required a total of 10--15 accumulated man months of efforts during the development and first years of operation by a single experienced software engineer (Sliepen), and the present system can be modified and upgraded with small efforts. Added to this software development was the effort needed to design the DM and the development of the software for simulating its closed loop performance and for processing calibration data and calculating the various matrices that are used by the AO code. 

\section{Conclusions}
In this paper, we have described the concepts, design, implementation and calibration of the SST 85-electrode AO system, which delivers an image quality of unsurpassed quality. The AO system is also used for continuously monitoring and recording measurements of seeing (Fried's seeing parameter $r_0)$, as described by \citet{2019A&A...626A..55S}. This seeing information is important for the development of improved techniques for image reconstruction \citep[MFBD and MOMFBD techniques, see:][]{2002SPIE.4792..146L, 2005SoPh..228..191V}, and in particular for extensions of these techniques that allow statistical compensation for the degrading effects of high-order seeing aberrations that are neither compensated for by the AO system, nor by MOMFBD \citep{2010A&A...521A..68S,2022A&A...668A.129L}. 
This seeing information is also of relevance for the future European Solar Telescope \citep[EST;][]{2022A&A...666A..21Q}, which will be erected only about 65~m from SST.

As regards the question of what may be of relevance for the future development of solar and night-time AO systems, the following suggestions are made:
\begin{itemize}
\item The use of work station technology, operating under Linux, and with software written in C++, offers a cost effective approach to developing an AO system. The software of the present AO system required only 10-15 accumulated months of efforts. Future upgrades of the system, based on expected improvements of work station technology, likely will be possible with only small efforts
\item The use of the absolute difference square (ADF$^2$) algorithm for cross correlations is of particular interest because of the possibility of implementation of the absolute difference algorithm using MMX multimedia instructions. The presently used PSADBW instructions offered a factor 9 times faster implementation than a corresponding C-implementation. Note that the use of the ADF$^2$ algorithm also on pinhole images for calibrating the DM response matrix were made with noise levels of only 0\farcs018, suggesting the versatility of this algorithm for a wide range of targets
\item Based on previous developments of correlation trackers with increasingly large FOV, and the present AO system, the use of large FOV for cross correlations makes solar AO systems more robust in bad seeing conditions, and helps to average out the effects of high-altitude differential seeing effects
\item The use of a second microlens array with many more subapertures than used in closed loop, offers the possibility to improve the characterization of the electrode responses by calibrating the response matrix with high pupil resolution, but requires a sufficiently large FOV to accommodate the associated image displacements during calibration
\item The use of monomorph DMs with Shack Hartmann wavefront sensors may be a rewarding approach to optimising AO performance in the visible on meter-class telescopes and on much larger telescopes at NIR and IR wavelengths
\item Evaluating performance of solar AO systems is difficult but obviously important - we have demonstrated that this can be done by correlating measurements of Fried's $r_0$ parameter with measurements of granulation contrast \citep{2019A&A...626A..55S}.
\end{itemize}

Simulations (Sect. \ref{sect:electrode_layout}) suggest that a Strehl approaching 0.5 should be possible at 500 nm wavelength when $r_0$=7~cm. From Fig. 6, in \citet{2019A&A...626A..55S}, the panels for 485~nm and 525~nm (blue dots), we obtain a mean granulation contrast of about 7.1\% resp. 6.5\% when $r_0$=7~cm whereas the theoretical granulation contrast is 22\% resp. 19.6\%. This corresponds to a Strehl of about 0.32-0.33, suggesting that the present AO system is operating at a relatively high efficiency, as defined by \citet{1998PASP..110..837R}, of about 50\%, but data at other wavelengths and $r_0$ values \citep{2019A&A...626A..55S}, suggest a lower efficiency and the need for an additional (seeing unrelated) wavefront error of about 48~nm RMS to explain the data. We believe that further evaluation and improvement of the efficiency of the existing 85-electrode AO system may be a more rewarding choice for SST rather than developing a new higher order AO system - a similar conclusion was drawn in a more general context of night-time AO systems by \citet{1998PASP..110..837R}.

\begin{acknowledgements}
  The Swedish 1-m Solar Telescope is operated on the island of La
  Palma by the Institute for Solar Physics of Stockholm University in
  the Spanish Observatorio del Roque de los Muchachos of the Instituto
  de Astrof\'isica de Canarias. The Institute for Solar Physics is
  supported by a grant for research infrastructures of national
  importance from the Swedish Research Council (registration number
  2017-00625). This work was supported by the Swedish Research Council, grants number 822-2007-3215 and 621-2014-5738, and has received funding from the European Union’s Horizon 2020 research and innovation programme under grant agreement No 824135. Francois Rigaut is acknowledged for several valuable discussions and suggestions relating to the design of bimorph deformable mirrors. Marcos van Dam is thanked for verifying the performance of the 85-electrode DM design, and Tomas Hillberg is acknowledged for checks and valuable comments on the AO software. Manolo Collados and Dan Kiselman are thanked for valuable discussions on the daytime seeing on La Palma.
\end{acknowledgements}

\appendix

\section{Control modes}
\label{sec:control-modes}

The family of 85 electrode influence functions, $\{E_i\}$, shown in
Fig.~\ref{fig:DM_response}, define a function space of possible mirror
shapes. To optimally control the mirror, we combine these functions
into a set of basis modes that are both orthonormal (with respect to
an inner product over the pupil) and statistically independent for
Kolmogorov turbulence. \citet{2000SPIE.4007..620L} describe a double
diagonalization procedure\footnote{``Diagonalization'' is a common
  term for finding the eigenmodes. A matrix $\mathbf{A}$ can be
  written as $\mathbf{A} = \mathbf{M}^\text{T} \mathbf{V} \mathbf{M}$,
  where $\mathbf{V}$ is the diagonal matrix that contains the
  eigenvalues and $\mathbf{M}$ are the eigenmodes.} that accomplishes
this, and also show that the resulting modes have the required
properties. In this appendix we follow their procedure, using their
notation. Symbols used here may therefore have a different meaning
than in the main text, unless explicitly stated otherwise.

Mathematically, the basis we need, $\cmmatrix$, is defined by the
requirements that its geometrical correlation matrix is the identity
matrix, and that its atmospheric covariance matrix is diagonal. The
requirements can be written as a system of matrix equations,
\begin{align}
  \cmmatrix^{-1}\cdot
  \acmatrix\cdot
  (\cmmatrix^\text{T})^{-1} &= \mathbf{\Sigma},  \label{eq:1}\\ 
  \cmmatrix^\text{T}\cdot \gcmatrix\cdot \cmmatrix &= \mathbf{I}. \label{eq:2}
\end{align}
Here $\acmatrix$ is the atmospheric covariance matrix and
$\gcmatrix$ is the geometrical correlation matrix of the
influence functions. $\mathbf{\Sigma}$ is a diagonal matrix of control
mode variances and $\mathbf{I}$ is the identity matrix expressing the
orthonormality of the control modes.


We first need to calculate the matrices $\acmatrix$ and
$\gcmatrix$, which is quite simple because we have already
expressed all the influence functions as sums of KL functions, $K_k$,
so that
\begin{equation}
  E_i(r,\varphi) = \sum_k G_{k,i} K_k(r,\varphi),
  \label{eq:6}
\end{equation}
where the coefficients $G_{k,i}$ define the matrix $\ifmatrix$ (referred to as $\ifmatrix_{\text{h}}$ in Sect.~\ref{sec:establ-contr-matr} when measured with the high-resolution WFS).

The elements of the geometrical correlation matrix, $\gcmatrix$,
are the inner products of the influence functions,
\begin{equation}
  \label{eq:7}
  \begin{split}
    \Delta_{ij} &= \ip{E_i}{E_j} 
    = \bigl\{ \text{Eq.~(\ref{eq:6})} \bigr\} \\
    & = \ip{\sum_k \ifelement_{k,i} K_k}{\sum_{k'} \ifelement_{k',j} K_{k'}}\\
    & = \sum_k\sum_{k'}\ifelement_{k,i} \ifelement_{k',j} 
    \ip{K_k}{K_{k'}} \\
    & = \sum_k \ifelement_{k,i} \ifelement_{k,j},
  \end{split}
\end{equation}
where we used the orthonormality of the KL functions,
$\ip{K_k}{K_{k'}} = \delta_{k,k'}$. In matrix form, we can write
Eq.~(\ref{eq:7}) as
\begin{equation}
  \label{eq:8}
  \gcmatrix = \ifmatrix^\text{T}\ifmatrix.
\end{equation}

Similarly, the elements of the statistical covariance matrix
$\acmatrix$ are the atmospheric covariances of the influence
functions,
\begin{equation}
  \label{eq:12}
  \begin{split}
    \acelement_{i,j} &= \cov(E_i,E_j) 
    = \bigl\{ \text{Eq.~(\ref{eq:6})} \bigr\} \\
    &= \cov\left(
      \sum_k \ifelement_{k,i} K_k \,,
      \sum_{k'} \ifelement_{k',j} K_{k'}
    \right)\\
    & = \sum_k\sum_{k'}\ifelement_{k,i} \ifelement_{k',j} \cov(K_k,K_{k'})\\
    & = \sum_k\ifelement_{k,i} \ifelement_{k,j} \sigma_k^2,
  \end{split}
\end{equation}
where we used the fact that the KL functions are statistically
uncorrelated with variances $\sigma_k^2$, so
$\cov(K_k,K_{k'})=\delta_{kk'}\sigma_k^2$. In matrix form we can write
Eq.~(\ref{eq:12}) as
\begin{equation}
  \label{eq:13}
  \acmatrix = \ifmatrix^\text{T}\cdot \mathbf{S}\cdot \ifmatrix,
\end{equation}
where $\mathbf{S} = \diag\{\sigma_k^2\}$.

The first step in  the recipe of \citet[their appendix
A]{2000SPIE.4007..620L}  is to diagonalize the correlation matrix
$\gcmatrix$, which will make our modes orthonormal. We write
\begin{equation}
  \label{eq:10}
  (\mmmatrix')^\text{T}\cdot \gcmatrix\cdot \mmmatrix'
  = \esmatrix^2 \quad\Rightarrow\quad \gcmatrix = \mmmatrix'\cdot \esmatrix^2\cdot (\mmmatrix')^\text{T},
\end{equation}
where $\mmmatrix'$ is an orthogonal matrix and $\esmatrix^2$ is a
diagonal matrix. The solution to this is obtained with singular value
decomposition (SVD),
\begin{equation}
  \label{eq:19}
  \gcmatrix =  \mathbf{U}\cdot \mathbf{W}\cdot \mathbf{V}^\text{T},
\end{equation}
which gives us $\esmatrix^2$ and the orthogonal basis $\mmmatrix'$
by identification with Eq.~(\ref{eq:10}). We make an
orthonormal basis by dividing with the eigenvalues,
\begin{equation}
  \label{eq:16}
  \mmmatrix = \mmmatrix'\cdot \esmatrix^{-1}.  
\end{equation}
The columns of $\mmmatrix$ are the voltages needed to make these
mirror modes.

We have so far not taken the atmospheric turbulence into account. We
proceed to do that by calculating the correlation matrix of the mirror
modes,
\begin{equation}
  \label{eq:14}
  \acmatrix' = \mmmatrix^{-1}\cdot \acmatrix\cdot (\mmmatrix^{-1})^\text{T},
\end{equation}
using the $\mmmatrix$ matrix from the previous paragraph. We then
diagonalize this matrix through identification with SVD as above, 
\begin{equation}
  \label{eq:17}
 \mathbf{\Sigma} = \dmmatrix^\text{T}\cdot \acmatrix'\cdot \dmmatrix,
\end{equation}
which yields $\dmmatrix$ and $\mathbf{\Sigma}$. The variances in $\mathbf{\Sigma}$ are plotted in Fig.~\ref{fig:DM_control2}.

The modal base matrix is then calculated as
\begin{equation}
  \label{eq:18}
  \cmmatrix = \mmmatrix\cdot\dmmatrix,
\end{equation}
which is an 85$\times$85 element matrix that provides a unique connection between the 85 electrodes of the DM and the 85 mode coefficients. This is used to calculate the voltages $ \mathbf{v}$ needed to produce the control modes with amplitudes $\mathbf{m}$
\begin{equation}
  \mathbf{v} = \cmmatrix\cdot \mathbf{m}.
\end{equation}
Combining this with Eq.~\ref{eq:3} in Sect.~\ref{sec:establ-contr-matr}, we obtain
\begin{equation}
    \mathbf{k} = \ifmatrix\cdot \mathbf{v} = (\ifmatrix\cdot \cmmatrix)\cdot \mathbf{m}  = \boldsymbol{\alpha}\cdot \mathbf{m},
\end{equation}
such that the matrix $ \boldsymbol{\alpha}$ provides the KL coefficients from the control mode coefficients. The control modes $S_i(r,\varphi)$, shown in Fig.~\ref{fig:DM_control}, are then obtained as expansions in KL modes
\begin{equation}
  \label{eq:sysmode}
  S_i(r,\varphi) = \sum_k \alpha_{k,i}  K_k(r,\varphi).
\end{equation}

\end{document}
